\documentclass[a4paper,11pt]{article}
\pdfoutput=1 

\usepackage{jheppub} 
\usepackage{amsmath,amssymb,amsthm,amscd,graphicx,hyperref}
\hypersetup{colorlinks=true, urlcolor=blue}
\input epsf.sty

\theoremstyle{definition}

\newcommand{\CA}{{\cal A}}

\newcommand{\CN}{{\cal N}}
\newcommand{\CO}{{\cal O}}

\def\IZ{{\mathbb Z}}

\newcommand{\re}{{\rm e}}
\newcommand{\ri}{{\rm i}}
\newcommand{\rd}{{\rm d}}

\newcommand{\be}{\begin{equation}}
\newcommand{\ee}{\end{equation}}
\newcommand{\ba}{\begin{aligned}}
\newcommand{\ea}{\end{aligned}}
\newcommand{\ben}{\begin{eqnarray}\displaystyle}
\newcommand{\een}{\end{eqnarray}}

\newcommand{\sectiono}[1]{\section{#1}\setcounter{equation}{0}}

\newdimen\tableauside\tableauside=1.0ex
\newdimen\tableaurule\tableaurule=0.4pt
\newdimen\tableaustep
\def\phantomhrule#1{\hbox{\vbox to0pt{\hrule height\tableaurule width#1\vss}}}
\def\phantomvrule#1{\vbox{\hbox to0pt{\vrule width\tableaurule height#1\hss}}}
\def\sqr{\vbox{%
  \phantomhrule\tableaustep
  \hbox{\phantomvrule\tableaustep\kern\tableaustep\phantomvrule\tableaustep}%
  \hbox{\vbox{\phantomhrule\tableauside}\kern-\tableaurule}}}
\def\squares#1{\hbox{\count0=#1\noindent\loop\sqr
  \advance\count0 by-1 \ifnum\count0>0\repeat}}
\def\tableau#1{\vcenter{\offinterlineskip
  \tableaustep=\tableauside\advance\tableaustep by-\tableaurule
  \kern\normallineskip\hbox
    {\kern\normallineskip\vbox
      {\gettableau#1 0 }%
     \kern\normallineskip\kern\tableaurule}%
  \kern\normallineskip\kern\tableaurule}}
\def\gettableau#1{\ifnum#1=0\let\next=\null\else
\squares{#1}\let\next=\gettableau\fi\next}

\newcommand{\figref}[1]{Fig.~\protect\ref{#1}}

\title{\boldmath Holomorphic Anomaly and Quantum Mechanics}

\author{Santiago Codesido and Marcos Mari\~no}

\affiliation{D\'epartement de Physique Th\'eorique et Section de Math\'ematiques\\
Universit\'e de Gen\`eve, Gen\`eve, CH-1211 Switzerland}

\emailAdd{santiago.codesido@unige.ch, marcos.marino@unige.ch} 

\abstract{We show that the all-orders WKB periods of one-dimensional quantum mechanical oscillators are governed by the refined holomorphic anomaly equations 
of topological string theory. We analyze in detail the double-well potential and the cubic and quartic oscillators, 
and we calculate the WKB expansion of their quantum free energies by 
using the direct integration of the anomaly equations. We reproduce in this way all known 
results about the quantum periods of these models, which we express in terms of modular forms on the WKB curve. 
As an application of our results, we study the large order behavior of the WKB expansion in the case of the double well, which 
displays the double factorial growth typical of string theory. 
}

\begin{document}

\maketitle
\flushbottom

\sectiono{Introduction}
In recent years, and through different perspectives \cite{adkmv,ns,mm-stms}, it has become clear that topological strings on 
local Calabi--Yau geometries are closely related to quantum mechanical problems and to integrable systems. This has led to a fruitful interplay 
between topological string theory (and its close cousin, $\CN=2$ gauge theory in 4d) and spectral problems, 
in which the results in one field have shed unexpected light on the other. For example, the so-called Nekrasov--Shatashvili (NS) limit of $\CN=2$ gauge theories 
is equivalent to the all-orders WKB quantization of certain quantum integrable systems. By using the Nekrasov instanton partition function \cite{nekrasov},  
one can then obtain new, exact quantization conditions for these systems. These include simple 
quantum mechanical models, like the modified Mathieu equation 
(see for example \cite{hm} for a detailed study 
of this case). Studies of the connections between Quantum Mechanics and refined topological string/gauge theory have been made in 
for example \cite{mirmor,acdkv, he,krefl,krefl2,bd,kpt,ashok}. 

In this paper, we suggest a more general implication of topological string theory for Quantum Mechanics. We claim that the 
refined holomorphic anomaly equations of \cite{kw,hk}, characterizing the refined topological string free energies, govern the all-order 
WKB periods of generic one-dimensional quantum mechanical systems. 

Let us recall that the holomorphic anomaly equations were first introduced in \cite{bcov} to describe 
the non-holomorphic dependence of standard topological string amplitudes. It turns out 
that these equations determine these amplitudes to a large extent. In some local CY geometries, 
the holomorphic anomaly equations, combined with modularity 
and appropriate boundary conditions, can be used to compute recursively (and efficiently) the 
topological string free energies at all genera \cite{hk-ha, hkr}. 
Since topological strings are closely related to many other physical systems, the holomorphic anomaly 
equations have provided very powerful tools to study 
a variety of problems. For example, the large $N$ expansion of matrix models is also governed 
by the holomorphic anomaly \cite{dv,hk-ha,emo}. This has made 
possible to obtain the all-genus free energy of some matrix models and/or gauge theories 
(see \cite{kmr,dmp} for some relevant examples.) 

The holomorphic anomaly equations of \cite{bcov} can be extended to refined topological string theory and to $\CN=2$ gauge theories \cite{kw,hk}.  
As we noted above, the NS limit of certain gauge theories is closely related to quantum mechanical problems, like the modified 
Mathieu equation. Therefore, one can 
use the holomorphic anomaly equations to analyze the all-orders WKB expansion of these problems. 
In this paper we argue that this connection 
is more general, i.e. we argue that, given a general quantum mechanical problem in one dimension, not necessarily 
related to supersymmetric gauge theories or topological strings, its 
all-orders WKB expansion is captured by the refined holomorphic anomaly equations in the NS limit.  
These equations can be in fact formulated for any algebraic curve obtained from the classical potential: they 
only require the special geometry determined by the classical periods, and the first quantum correction to the free energy, which can be 
expressed in terms of the discriminant of the curve \cite{kw,hk}.  

In order to test our claim, we consider the most famous one-dimensional models in Quantum Mechanics: the cubic 
oscillator, the symmetric double-well, and the 
quartic oscillator. The modern, non-perturbative treatment of these models started with the 
work of Voros \cite{voros,voros-return} and Zinn--Justin \cite{zj} in the early 1980s. In those papers, 
a series of ``exact" quantization conditions were found, involving 
the Borel-resummed all-orders WKB periods of \cite{dunham}, as well as explicit non-perturbative corrections. 
These exact quantization conditions can be also 
derived in the framework of the uniform WKB method \cite{shcp,al-cas,al-cas-2,ahs,alvarez}. 
In this paper, we explicitly show that the refined holomorphic anomaly equations, combined 
with the direct integration method of \cite{hk-ha,gkmw}, determine the non-perturbative 
quantum period appearing in the quantization conditions, as a function of the perturbative quantum period. In other words, 
the refined holomorphic anomaly equations govern the all-orders 
WKB periods for one-dimensional oscillators, at least when the underlying geometry 
has genus one\footnote{In the case of the double well, the connection with the refined 
holomorphic anomaly could have been anticipated by using the relation to the cubic matrix model found in \cite{krefl}, 
together with the results of \cite{huang-beta}. However, our claim is more general and does not depend on a matrix model 
intermediary. For example, the cubic oscillator does not have any known realization as a $\beta$-deformed matrix model, 
yet it satisfies the refined holomorphic anomaly equation, as we show in this paper.}. 
As a spinoff, we calculate the quantum free energies of these models explicitly and recursively in terms of modular forms on the WKB curve. 
This provides a new and efficient method to calculate the all-orders WKB expansion of the non-perturbative $A$ function appearing in the 
quantization conditions of \cite{zj,zjj1,zjj2,jsz}, in terms of modular forms. 

As a simple application of our results, we study the large order behavior of the WKB expansion for the energy of the symmetric double well. The 
standard perturbative series in the coupling constant for the ground state energy has been analyzed in much detail in for example \cite{bpz,zjj1,zjj2}. This is a numerical 
series whose coefficients grow like $n!$, and its large order behavior is controlled by the two-instanton correction. 
The WKB expansion, in contrast, is a series in $\hbar^2$. Its coefficients are functions of the modulus $\nu$ (which encodes the 
quantum number), and they grow like $(2n)!$. This is of course the typical growth of the genus expansion in string theory and of the 
$1/N$ expansion \cite{shenker,mm-rev,mm-book}. The large order behavior of this series is still controlled by the two-instanton correction, and we derive and test a 
precise formula for the growth of the coefficients up to, and including, the term of order $1/n$. 

This paper is organized as follows. In section 2 we review some basic facts about the all-orders WKB method and the refined holomorphic 
anomaly equations, and we present our main conjecture connecting both approaches. We also give some details about the direct integration 
of the anomaly equations. Section 3 illustrates our general considerations with a detailed analysis of the cubic oscillator and the double well. 
The quartic oscillator can be also 
obtained by an $S$-duality transformation of the double-well. In section 4 we present the analysis of the large order behavior of the WKB expansion in the 
example of the double well. Finally, in section 5 we present our conclusions and list some open problems for future research.

\sectiono{The all-orders WKB method and the holomorphic anomaly}

 In this section we review the all-orders WKB method of Dunham \cite{dunham} (see for example \cite{gp2} for a clear textbook presentation). We start with the 
 Schr\"odinger equation for a one-dimensional particle in a potential $V(x)$, 
 \be
 \label{schrodinger}
\hbar^2 \psi''(x) +p^2(x) \psi(x)=0, \qquad p(x) ={\sqrt{2(\xi-V(x))}}, 
\ee
where $\xi$ denotes the energy of the particle, and we have set the mass $m=1$. The WKB ansatz for the wavefunction is 
\be
\label{yansatz}
\psi(x)=\exp \left[ {\ri \over \hbar} \int^x Q(x') \rd x' \right]. 
\ee
The Schr\"odinger equation for $\psi(x)$ becomes a Riccati equation for $Q(x)$, 
\be
\label{riccati}
 Q^2(x)-\ri \hbar\frac{\rd Q(x)}{\rd x}=p^2(x), 
\end{equation} 
which can be solved in power series in $\hbar$:
\be
\label{yps}
Q(x)=\sum_{k=0}^\infty Q_k(x)\hbar^k. 
\ee
The functions $Q_k(x)$ can be computed recursively as 
\begin{equation}
\label{y-recur}
\ba
 Q_0(x)&=p(x), \\
  Q_{n+1}(x)&=\frac{1}{2Q_0(x)}\left(\ri\frac{\rd Q_n(x)}{\rd x}-\sum_{k=1}^n Q_k(x)Q_{n+1-k}(x)\right).
  \ea
 \ee
 If we split the formal power series in (\ref{yps}) into even and odd powers of $\hbar$, 
 \be
 \label{yyp}
 Q(x) = Q_{\rm odd} (x) + P(x),
\ee
one finds that $Q_{\rm odd}(x)$ is a total derivative, 
\be
Q_{\rm odd}(x)={\ri \hbar \over 2} {P'(x) \over P(x)}={\ri \hbar \over 2} {\rd \over \rd x} \log  P(x).
\ee
It follows that only $Q_1$ contributes to period integrals of $Q(x)$. In addition, the wavefunction (\ref{yansatz}) can be written as 
\be
\psi(x) ={1\over {\sqrt{ P(x)}}} \exp \left[ {\ri \over \hbar} \int^x P(x') \rd x' \right].
\ee

Let us now consider the Riemann surface $\Sigma$ defined by 
\be
\label{alg-curve}
y^2= p^2(x). 
\ee
The turning points of the WKB problem are the points where $p^2(x)=0$, and correspond to the branch points of the curve (\ref{alg-curve}). 
We will assume that turning points are simple. We will also restrict ourselves to curves of genus one, in which there are only two different, independent one-cycles 
$A$ and $B$ encircling turning points. The $A$-cycle corresponds to an allowed region for the 
classical motion, while the $B$-cycle corresponds to a forbidden region. We can now consider the periods of the one-form $y(x) \rd x$ along these cycles, 
\be
t= {1\over 2 \pi} \oint_A y (x) \rd x, \qquad t_D= -\ri \oint_B y(x) \rd x, 
\ee
with appropriate choices of branch cuts for the function $y(x)$. We will refer to these as {\it classical periods}. They depend on the energy 
$\xi$ and on the parameters of the potential. We note that the $A$-cycle corresponds 
to the ``perturbative" cycle, while the $B$-cycle corresponds 
to the ``tunneling cycle", and it typically encodes non-perturbative corrections. As in (local) mirror symmetry, it is useful to introduce a prepotential or 
classical free energy $F_0(t)$ by the equation 
\be
\label{prepot}
t_D ={\partial F_0 \over \partial t}. 
\ee
We should regard $t$ as a flat coordinate parametrizing the complex structure of the curve (\ref{alg-curve}). The inverse function expressing $\xi$ as a function of the classical $A$-period, 
which we will denote as $\xi_0(\nu)$, 
will be called the {\it classical mirror map}. This function is nothing but the classical Birkhoff series expressing the energy $\xi$ as a function of the classical action $\nu$ 
(see \cite{ags,alvarez3}). 

The all-orders WKB method allows one to define ``quantum" versions of the classical periods, by simply using the formal power series for $P(x)$, 
\be
P(x) =\sum_{n \ge 0} Q_{2n}(x) \hbar^{2n}. 
\ee
In this way, we can promote the classical periods to ``quantum" periods, 
\be
\nu =  {1\over 2 \pi \ri} \oint_A P (x) \rd x, \qquad {\partial F \over 
\partial \nu}=-\ri  \oint_B P(x) \rd x. 
\ee
Note that both of them are defined by formal power series expansions in $\hbar^2$, 
\be
\nu= \sum_{n \ge0} t^{(n)} \hbar^{2n}, \qquad  {\partial F\over 
\partial \nu}= \sum_{n \ge 0} t_D^{(n)} \hbar^{2n}, 
\ee
and the leading order term of this expansion, which is obtained as $\hbar \rightarrow 0$, gives back the classical periods. 

The quantum $A$-period 
defines what we will call, following \cite{acdkv}, the {\it quantum mirror map}:
\be
\label{qmm}
\xi(\nu)= \sum_{n \ge 0} \xi_n(\nu) \hbar^{2n}. 
\ee
Note that the first term of (\ref{qmm}), $\xi_0(\nu)$, is the classical 
mirror map. In the context of one-dimensional quantum mechanical oscillators, with a coupling constant $g$, 
the quantum mirror map was studied in detail in \cite{ags,alvarez3}, as a formal power 
series of quantum corrections to the classical Birkhoff series $\xi_0(\nu)$. It is closely related to the 
stationary perturbation series for the energy. Indeed, let us set
\be
\label{num}
\nu= \hbar \left( m+{1\over 2} \right), 
\ee
where $m$ is the quantum number for the energy level. If we now expand each $\xi_n(\nu)$ as a power series in the coupling constant $g$, and we rearrange 
the terms in (\ref{qmm}) in powers of $g$, we recover from $\xi(\nu)$ the Rayleigh--Schr\"odinger series. 

The quantum $B$-period defines the {\it quantum free energy} as function 
of the full quantum period $\nu$, 
\be
F(\nu)= \sum_{n\ge0 } F_n(\nu) \hbar^{2n}. 
\ee
This quantity is the analog of the NS free energy in supersymmetric gauge theories and topological strings. It is then 
natural to conjecture that it is governed by the same set of equations, namely, the refined holomorphic anomaly equations of \cite{kw,hk}
(see \cite{hkk,huang} for further developments.) Let us briefly review these equations. 

We consider the B-model refined topological string on a local Calabi--Yau manifold described by a Riemann surface $\Sigma$. The basic 
perturbative quantities are the free energies $F^{(g_1, g_2)}(t_i)$, with $g_1, g_2 \ge 0$ (see for example \cite{hk,ckk} for a summary and 
a list of references). The standard topological string free energy corresponds to $g_1=0$, while the NS limit corresponds to $g_2=0$. We will denote 
\be
F^{(n,0)}= F_n.  
\ee
Note that $F_g$ usually denotes the standard topological string free energies $F^{(0,g)}$, but these will not play any r\^ole in this paper. 
The refined free energies can be promoted to non-holomorphic functions of the moduli $t_i$ and their 
complex conjugates $\bar t_i$, $F^{(g_1, g_2)}(t_i, \bar t_i)$. The refined holomorphic anomaly equations 
govern the anti-holomorphic dependence of these functions, and they read, 
\be
\label{rha}
{\partial F^{(g_1, g_2)} \over \partial \bar t_k}= {1\over 2 \gamma} \overline{C}_{\bar k}^{lm} \left( D_l D_m F^{(g_1, g_2-1)} 
+ \sum_{0<r_1+r_2< g_1+g_2} D_l F^{(r_1, r_2)} D_m F^{(g_1-r_1, g_2-r_2)}\right). 
\ee
We have introduced a constant $\gamma$ to take into account different normalizations for the free energies. These equations are valid for $g_1+g_2\ge 2$. 
They involve the metric $G_{k \bar m}$ on the moduli space of complex structures, 
as well as the corresponding covariant derivatives $D_i$. $C_{ijk}$ is the Yukawa coupling 
\be
\label{yukawa}
C_{ijk}={\partial^3 F_0 \over \partial t_i \partial t_j \partial t_k}, 
\ee
and
\be
\overline{C}_{\bar k}^{lm}= G^{l \bar p} G^{m \bar n} \overline{C}_{\bar  p \bar n \bar k}. 
\ee
If we denote by $\tau$ the period matrix of 
$\Sigma$, the metric is given by 
\be
G_{i \bar j}= -\ri \pi \left( \tau- \overline \tau \right)_{ij}. 
\ee
In the NS limit, the first term in the r.h.s. of (\ref{rha}) drops out, and we obtain the simplified equations, 
\be
\label{rha-ns}
{\partial F_n \over \partial \bar t_k}= {1\over 2 \gamma} \overline{C}_{\bar k}^{lm}  \sum_{r=1}^{n-1} D_l F_r D_m F_{n-r},  \qquad n\ge 2. 
\ee
The equations (\ref{rha-ns}) have to be supplemented with an explicit expression for $F_1$. It turns out that \cite{hk,kw}
\be
\label{f1}
F_1= -{1\over \alpha} \log \Delta, 
\ee
where $\Delta$ is the discriminant of the curve $\Sigma$, and $\alpha$ is a real number (in the examples considered in this paper, $\alpha=24$). 
Using (\ref{f1}) as the initial datum, as well as the special geometry of the 
moduli space, the holomorphic anomaly equation (\ref{rha-ns}) determines the functions $F_n$ 
recursively, up to a purely holomorphic dependence on the moduli which is usually called the {\it holomorphic ambiguity}. 

It is very important to note that, in the holomorphic anomaly equations, the functions $F_n$ are obtained as functions of the moduli of the curve, and in order to 
express them in terms of the period $\nu$ one uses the {\it classical} mirror map $\xi_0(\nu)$, i.e. the inverse of the {\it classical} period $t=t(\xi)$.  

It was noted in \cite{hk-ha,gkmw} and subsequent works that the holomorphic anomaly equation can be solved in a very efficient way by using the method of 
``direct integration." In this method, one exploits the symmetries of the problem and re-expresses all quantities in terms of modular forms of the period matrix $\tau$. The method 
is most efficient for elliptic curves, although it can be successfully used as well for curves of higher genus \cite{kpsw}. For genus one curves, 
with a single geometric modulus, there is a single Yukawa coupling 
\be
Y=C_{ttt}.
\ee
 The non-holomorphic dependence occurs through the so-called 
propagator $S^{tt}$, which is defined by 
\be
\overline{C}^{tt}_{\bar t}=\bar{\partial}_{\bar t}S^{tt}.  
\ee
In all known models with curves of genus one, the propagator can be written as 
\be
S^{tt} =- {1\over 3 \beta} \widehat E_2(\tau, \bar \tau), 
\ee
where $\beta$ is an appropriate proportionality constant, $\widehat E_2(\tau, \bar \tau)$ is the non-holomorphic modular form 
\be
\widehat E_2(\tau, \bar \tau)= E_2(\tau)- {3 \over \pi {\rm Im}\, \tau}, 
\ee
and $E_2(\tau)$ is the weight two Eisenstein series. 
The elliptic modulus $\tau$ is related to the prepotential by the equation, 
\be
\tau ={\beta  \over 2 \pi \ri} {\partial^2 F_0 \over \partial t^2}.  
\ee
The covariant derivative $D_t$ can then be written as
\be
D_t=  \beta Y D_\tau, 
\ee
where $D_{\tau}$ is the Maass derivative acting on (almost-holomorphic) modular forms of weight $k$:
\be
D_\tau=\frac{1}{2\pi\ri}\frac{\rd}{\rd\tau}-\frac{k}{4\pi\text{Im} \tau}.
\ee
Putting all these ingredients together, one can write the anomaly equation in the form 
\be
\frac{\partial F_n}{\partial\widehat{E}_2}=-{ \beta \over  6 \gamma} 
Y^2\sum_{r=1}^{n-1}D_{\tau}F_r D_\tau F_{n-r},\qquad n \ge 2.
\ee
This can be integrated as 
\be
\label{fncn}
F_n(\tau, \bar \tau)= \sum_{\ell=1}^{2n-3} c^{(n)}_{\ell}(\tau) \widehat E^\ell_2(\tau, \bar \tau) + f_n(\tau). 
\ee
The free energies $F_n(\tau, \bar \tau)$ are non-holomorphic modular forms of weight zero w.r.t. the appropriate monodromy group of the 
elliptic curve $\Sigma$. Therefore, in (\ref{fncn}), $c^{(n)}_\ell(\tau)$ are 
modular forms of weight $-2 \ell$, which are completely fixed by the holomorphic anomaly equations. The original holomorphic 
amplitudes are obtained by simply taking $\bar \tau \rightarrow \infty$, keeping $\tau$ fixed. This amounts to replacing 
\be
\widehat  E_2(\tau, \bar \tau) \rightarrow E_2(\tau)
\ee
in (\ref{fncn}). 

The function $f_n(\tau)$ appearing in (\ref{fncn}) is the holomorphic ambiguity. In order to fix it, 
one first finds an appropriate parametrization for it in terms 
of modular forms. This parametrization is determined by the monodromy group of $\Sigma$, and by the structure of the Yukawa coupling. 
Once this parametrization has been found, fixing the ambiguity reduces to determining a finite set of coefficients at each order $n$. To calculate these, 
we have to impose boundary conditions at special points in the moduli space. In the examples involving quantum mechanical models,
we will look at singular points of the curve, i.e. at points where the discriminant of the curve vanishes. 
One of these points, which we will call the {\it conifold} point, corresponds to $\nu=0$. Expanding around 
$\nu=0$ is equivalent to performing a perturbative expansion in the coupling constant. 
Near this point, the holomorphic free energies $F_n(\nu)$ are the sum of a singular part 
and of a regular part. We will write, 
\be
\label{sr}
F_n(\nu) =F_n^{\rm s} (\nu) + F_n^{\rm r}(\nu). 
\ee
The singular part has the following structure. For $n=0,1$, it is of the form,
\be
\label{singfn}
\ba
F^{\rm s}_0(\nu)&=k_0 {\nu^2 \over 2} \left( \log\left( {\nu \over \nu_0} \right) - {3 \over 2} \right),\\
F^{\rm s}_1(\nu)&=-{k_1 \over 24} \log(\nu), 
\ea
\ee
while for $n\ge 2$ there is a pole or order $2n-2$, 
\be
\label{coni-bound}
F_n^{\rm s}(\nu) =k_n  {(1-2^{1-2n}) (2n-3)! B_{2n} 
\over (2n)!}\nu^{2-2n},  \qquad n \ge 2.
\ee
In (\ref{singfn}) and (\ref{coni-bound}), $k_n$ is of the form $r_1 r_2^n$, where $r_{1,2}$ are constants, 
and takes into account possible overall normalizations of the free energies and the parameter $\nu$. We will fix these normalizations, and the value 
of $k_n$, in a case by case basis in the examples below. In the first line of (\ref{singfn}), $\nu_0$ is a constant which depends on the model. In (\ref{coni-bound}), 
$B_{2n}$ are Bernoulli numbers. On the other hand, 
\be
\label{freg}
F_n^{\rm r}(\nu)=\CO(1), \qquad  n \ge 0, 
\ee
 is regular at $\nu=0$. The behavior (\ref{coni-bound}), (\ref{freg}) is called the {\it gap condition} at the conifold point.  
It was first discovered in \cite{hk-ha}, in the study of standard topological string amplitudes and of gravitational corrections to Seiberg--Witten theory. 
It was generalized in \cite{hk,hkk,ckk} to the refined case. The behavior 
 (\ref{coni-bound}) imposes $2n-2$ conditions on the holomorphic ambiguity, but it does not fix it completely. 
 Additional conditions are obtained by looking at other singular points of the curve. There is typically a dual conifold point, where a dual period $\nu_D$, 
 obtained by a symplectic trasnsformation of $\nu$, vanishes. One then looks at the quantum free energies in the 
 symplectic frame appropriate for the dual conifold point, $F_n^D$, and studies their behavior near $\nu_D=0$. It turns out that the 
 dual free energies satisfy a dual gap condition, of the form 
\be
\label{d-coni-bound}
F^D_n(\nu_D) = k^D_n {(1-2^{1-2n}) (2n-3)! B_{2n} 
\over (2n)!}\nu_D^{2-2n}+ \CO(1), 
\ee
where $k_n^D$ is another normalization constant. Very often, 
the combination of (\ref{coni-bound}) and (\ref{d-coni-bound}) fixes completely the holomorphic ambiguity. 
This will be also the case in models arising in one-dimensional quantum oscillators, where the gap condition (\ref{coni-bound}) turns out to be 
equivalent to a well-known regularity property of the all-orders WKB periods. 

In the next section we will set up and solve the anomaly equations (\ref{rha-ns}) for various curves $\Sigma$ 
arising in one-dimensional Quantum Mechanics, 
of the form (\ref{alg-curve}). This gives the quantum free energies $F_n (\nu)$, which we will write as quasi-modular forms on the curve $\Sigma$. In order to determine the 
two quantum periods appearing in the WKB method, we need, in addition to the quantum free energy, the relation between $\nu$ and $\xi$. It turns out that 
the quantum mirror map (\ref{qmm}) can be obtained from the quantum free energy by using what we will call the {\it PNP relations} between 
the perturbative and the tunneling cycles. The PNP relations for one-dimensional quantum mechanical oscillators 
were first observed by \'Alvarez and Casares in \cite{al-cas-2} for the cubic oscillator, generalizing a result in \cite{hoe} for the Stark effect. 
They were established empirically for the quartic oscillator and the symmetric double well in \cite{ahs} and \cite{alvarez}, respectively, and extended recently to 
the periodic cosine potential and other one-dimensional models in \cite{du}. Originally, the PNP relations for one-dimensional oscillators 
were formulated as follows. Let us consider the function
\be
\label{afr}
A(\nu)= {\rd F^{\rm r} \over \rd \nu}= \sum_{n \ge 0} {\rd F^{\rm r}_n \over \rd \nu},  
\ee
where we set $\hbar=1$. Then, the derivative of the quantum mirror map and the $A$ function are related as,  
\be
\label{pnp-or}
{\rd \xi \over \rd \nu}=-{g^2\over \sigma}\left( C \nu +g^2 {\partial A \over \partial g^2} \right),  
\ee
where $\sigma$, $C$ are constants, and $g$ is the coupling constant of the Quantum Mechanical problem. Since this equation relates the tunneling period to the perturbative period, it provides a link 
between perturbative and non-perturbative physics. Relations like (\ref{pnp-or}) are also well-known 
for the NS limit of supersymmetric gauge theories, where they allow to compute the quantum mirror map from the 
quantum free energy. Derivations of the relationship (\ref{pnp-or}) in some particular models can be found in \cite{gorsky,bd}. As we will see in examples, by using the full quantum free energy, 
one can integrate the PNP relation to obtain, 
\be
\xi(\nu)=-\varphi-{1\over \sigma} g^4 {\partial F(\nu) \over \partial g^2}, 
\ee
where $\varphi$ is an integration constant, independent of $\nu$ and $g$. Using this relation, 
together with our explicit expressions for $F_n(\nu)$, we find explicit formulae for 
the $\xi_n(\nu)$ in (\ref{qmm}) with $n\ge 1$ in terms of modular forms. 

We then conjecture that the all-orders WKB periods of one-dimensional quantum systems satisfy the holomorphic anomaly equations (\ref{rha-ns}), and in the next sections we will 
test this conjecture in some non-trivial examples. It would be of course interesting to prove this conjecture rigorously. One possible strategy would be to adapt to our context recent 
results of Alba Grassi for topological strings in \cite{alba}, one of which is the following. Let us assume that the quantum periods transform as the 
classical periods under a symplectic transformation. Then, one can introduce a non-holomorphic dependence on the quantum free energy, in such a way 
that the resulting non-holomorphic objects satisfy (\ref{rha-ns}). Although these results were obtained in the context of topological string theory, they can be 
adapted to the quantum mechanical context of this paper. An important step in such a proof would then be to justify rigorously, in the quantum mechanical case, 
the assumptions of \cite{alba} concerning the transformation properties of the quantum periods. This would follow, in turn, if one could express the quantum periods as differential operators 
acting on the classical periods. This can be shown to be the case in many examples and for low orders of the $\hbar$ expansion (see e. g. the calculations in related 
examples in \cite{huang}), but a general proof is lacking. 

\sectiono{Examples}

\subsection{The cubic oscillator}

\begin{figure}[tb]
\begin{center}
\resizebox{75mm}{!}{\includegraphics{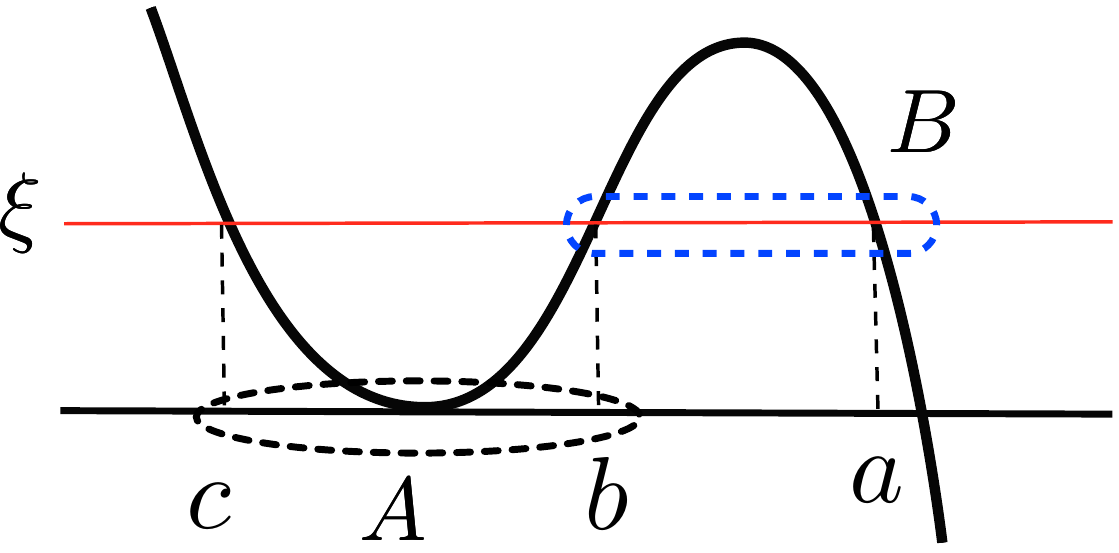}}
\end{center}
  \caption{The cubic oscillator. 
}
\label{cub-fig}
\end{figure}
The cubic oscillator is a one-dimensional quantum particle in the cubic potential
\be
V(x)={x^2\over 2} -g x^3. 
\ee
This is a very much studied example in Quantum Mechanics (see for example \cite{gp2} for a textbook presentation). This potential does not 
support bound states, but by imposing Gamow--Siegert boundary conditions, one finds an infinite tower of resonances which can be computed numerically 
by using complex dilatation techniques \cite{bender,alvarez-cub}. The resonant energies 
can be also obtained by Borel resummation of the stationary perturbative series in the coupling $g$ \cite{gg,alvarez-cub, al-cas-2,caliceti}. 
In the all-orders WKB approach, one can derive an 
exact quantization condition involving two different periods. 
To write down this condition, we denote the turning points of the potential by 
\be
c<b<a, 
\ee
as shown in \figref{cub-fig}. The $A$-cycle goes around the points $c$ and $b$, while the $B$-cycle goes around the points $b$ and $a$. The exact quantization 
condition can be obtained in a simple way by using the Voros--Silverstone connection formulae \cite{voros,voros-return,silverstone} 
(this is essentially the method followed in \cite{ddp}). Alternatively, 
it can be derived from the uniform WKB method \cite{al-cas,al-cas-2}. Since the resulting series is not Borel 
summable, there are two different quantization conditions, depending on the choice of lateral resummation. One choice gives 
a perturbative quantization condition, 
\be
\oint_A P(x) \rd x = 2\pi \hbar \left(n+{1\over 2}\right),  \qquad n \in \IZ_{\ge 0}, 
\ee
while the other choice gives 
  \be
  \label{qc-cubic}
 1+ \exp \left(  {\ri \over \hbar} \oint_A P(x) \rd x \right) + \exp\left( {\ri\over \hbar} \oint_B P(x) \rd x\right) =0. 
  \ee
  This quantization condition reduces to the previous one when the ``tunneling" integral over the $B$ cycle is neglected. The tunneling cycle gives a non-perturbative, 
 exponentially small correction to the perturbative result. The perturbative and the tunneling periods, regarded as quantum 
 $A$ and $B$ periods, define a quantum free energy $F(\nu)$ as 
 \be
 {\partial F(\nu) \over \partial \nu}=-\ri \oint_B P(x) \rd x. 
 \ee
The Voros multiplier associated to the tunneling period, and defined as 
 \be
 f(\nu) =\exp\left( {\ri \over \hbar} \oint_B P(x) \rd x\right),
 \ee
 has the following 
 structure (we set $\hbar=1$)
 \be
 \label{fnu}
 f(\nu)= {{\sqrt{2 \pi}} \over \Gamma\left( {1\over 2} + \nu\right)} \left( {8 \over g^2} \right)^\nu \re^{-A(\nu)}. 
 \ee
Here, $A(\nu)$ is the function defined in (\ref{afr}) and is given by the power series in $g$, 
 \be
 \label{anu-cub}
 A(\nu)={2\over 15 g^2} + \sum_{k =1}^ \infty 
 c^{(k)} (\nu) g^{2k}.  
 \ee
This structure can be obtained by using the uniform WKB method, as in \cite{al-cas, al-cas-2}, or the techniques of \cite{ddp}. 
The coefficients $c^{(k)}$ in (\ref{anu-cub}) turn out to be polynomials in $\nu$ with rational coefficients. They can be calculated explicitly by using 
 the asymptotic matching of WKB expansions involving Airy and parabolic cylinder functions, but their calculation is involved and time-consuming. 
 At the very first orders, one finds
 \be
 \label{c-cub}
 \ba
 c^{(1)}(\nu)&= {141 \nu^2\over 8} + {77 \over 32},\\
 c^{(2)}(\nu)&={7717  \nu^3\over 32} + {13937 \nu\over 128} , \\
 c^{(3)}(\nu)&=\frac{2663129 \nu ^4}{512}+\frac{5153379 \nu ^2}{1024}+\frac{43147783}{122880}. 
 \ea
 \ee
 A more comprehensive list can be found in \cite{casares-tesis}. One advantage of the method 
 developed in \cite{al-cas,al-cas-2} is that the tunneling period is expressed directly in terms of the quantum period $\nu$. 
 The tunneling period was also studied in \cite{jsz}, following 
 the ideas developed previously in \cite{zj,zjj1,zjj2}. In \cite{jsz}, the function $A(\nu)$ defined in (\ref{anu-cub}) was computed as a function of 
 the energy $\xi$, rather than the quantum period\footnote{The $A$ function of the cubic oscillator 
 obtained in \cite{jsz} was reexpressed in terms of $\nu$ in \cite{gt}, and the result can be seen to agree with the results 
 obtained previously by \'Alvarez and Casares in \cite{al-cas,al-cas-2}.}. 
 The calculation of $A(\nu)$ in \cite{jsz} is based on a direct evaluation of the period integrals, order by order in 
 $\hbar$, and it is quite involved\footnote{The convoluted integration method developed in \cite{zjj1,zjj2,jsz} to calculate the periods and their quantum corrections is 
not really needed in the case of curves of genus one. All the 
 integrals appearing in the cubic oscillator and other genus one examples can be computed in closed form in terms of elliptic integrals of the first and second kind.}. 
 We will see in this section that the holomorphic anomaly equation gives an extremely powerful method to compute the function $A(\nu)$ at arbitrary order, in terms of 
 modular forms.

The elliptic curve underlying the cubic oscillator is given by 
\be
\label{yx-cubic}
y^2=2 \xi - x^2 + 2 g x^3. 
\ee
It is convenient to rescale the variables so as to absorb the $g$ dependence:
\be
\label{rescaling}
x \rightarrow x/g , \quad y \rightarrow  y/g, \quad \xi \rightarrow  \xi/g^2,
\ee
and the new curve reads
\be
y^2=2 \xi - x^2 + 2  x^3. 
\ee
%
%
%
The classical periods can be written as 
\be
t= {1\over \pi} \int_c^b y(x) \rd x, \qquad t_D= -2 \ri \int_b^a y(x) \rd x. 
\ee
One can easily check that they satisfy the Picard--Fuchs equation 
\begin{equation}
		\left[\frac{15}{2} + \Delta(\xi)  \frac{\partial^2}{\partial \xi^2} \right] \Pi= 0, 
	\end{equation}
where
\be
\label{disc-cub}
\Delta(\xi)=\xi\left(54 \xi-1\right) 
\ee
is the discriminant of the elliptic curve (\ref{yx-cubic}) for $g=1$. 
From the Picard--Fuchs equation, we find that the period $t$ has an expansion around $\xi=0$ of the form,  
\be
\label{tex}
t = \sum_{n=1}^{\infty} c_n \xi^n,
\ee
where the coefficients satisfy the recursion relation, 
\be
c_n = \frac{3 (6 n-11) (6 n-7)}{2 (n-1) n} c_{n-1}, \qquad n \ge 2, \qquad c_1=1. 
\ee
It is also possible to write $t$ and $t_D$ in terms of elliptic integrals of the first and second kind:
\be
\ba
t&= {2 {\sqrt{2}} \over \pi} { (b-c)^2 {\sqrt{a-c}} \over 15 k^4} \left( (k')^2 (k^2-2) K(k) + 2 (k^4 + (k')^2) E(k) \right),\\
t_D &= 4\sqrt{2} \frac{(a-b)^2 (b-c)}{\sqrt{a-c}} \frac{2 (1+k^2 (k^2-1)) E(k') -k^2(1+k^2)K(k') }{15 k^2(k^2-1)^2},
\ea
\ee
where the elliptic modulus $k$ and its complementary $k'$ are  given by 
\be
\label{kmod-cub}
k^2={b-c \over a-c}, \qquad (k')^2= 1- k^2 ={a-b \over a-c}. 
\ee
An equivalent expression for the classical $A$ period in terms of hypergeometric functions was obtained in \cite{alvarez3}. 
In addition, the derivatives of the periods can be written in closed form as, 
\be
{\partial t \over \partial \xi}={1\over  \pi}{\sqrt{2 \over a-c}} K(k), \qquad {\partial t_D \over \partial \xi}=2{\sqrt{2 \over a-c}} K'(k), 
\ee
where as usual $K'(k)= K(k')$. Finally, the tau parameter is
\be
\label{tau-k}
\tau= \ri {K'(k) \over K(k)}= {1\over 2 \pi \ri} {\partial t_D \over \partial t}.  
\ee
The classical periods have the following expansion around $\xi=0$:
\be
\ba
t=& \xi + {15  \xi^2\over 4} + {1155  \xi^3 \over 16} +\frac{255255 \xi ^4}{128}+\cdots, \\
t_D=& {2\over 15 } +\xi  \left( \log\left({\xi  \over 8} \right)-1 \right)+{3  \xi^2 \over 8} \left( 10 \log\left({\xi  \over 8} \right)+47\right) + \cdots. 
\ea
\ee
Since an expansion in $\xi$ is an expansion in $g$, this makes contact with the perturbative series in $g$. Using these results and the definition (\ref{prepot}), 
we can compute the prepotential $F_0(t)$ in power series in $t$, around $t=0$, and one finds, 
\be
\label{prep-cubic}
F_0(t)= {t^2 \over 2} \left( \log\left( {t  \over 8}\right)- {3 \over 2} \right)+ {2t \over 15 } +{47  t^3\over 8} + {7717  t^4 \over 128}+\frac{2663129 t^5}{2560}+\cdots. 
\ee
Note that this has precisely the structure of a prepotential near a conifold point. 

By looking at the discriminant of the curve, we find that there are two singular points. The first one is the conifold point 
\be
\xi=0. 
\ee
The second one is the dual conifold point, 
\be
\xi={1\over 54}. 
\ee
This is the critical value in which the energy $\xi$ has the same magnitude as the height of the cubic barrier. It also corresponds to the 
value of the modulus $k^2=1$ in (\ref{kmod-cub}), so it sets the radius of convergence for the expansion (\ref{tex}). 
To formulate the dual problem, we introduce a dual energy
\be
\label{xid-xi}
\xi_D= {1\over 54}- \xi,  
\ee
such that $\xi_D=0$ is the dual conifold point. From the Picard--Fuchs equation, one finds that exchanging $\xi$ with $\xi_D$ 
corresponds to exchanging $t$ with $t_D$. In particular, the dual period $t_D$ has a regular expansion in terms of $\xi_D$:
\begin{equation}
   \frac{t_D}{2\pi} := \tilde 	t_D = \xi_D +\frac{15 \xi_D ^2}{4}+\frac{1155 \xi_D ^3}{16}+\frac{255255 \xi_D ^4}{128}+\CO\left(\xi_D ^5\right), 
	\end{equation}
	which is identical to the expansion of $t$ in terms of $\xi$. Physically, the dual situation corresponds to inverting the 
	cubic potential: the top of the barrier defines now the origin of energy, as in (\ref{xid-xi}), 
	and the tunneling and perturbative cycles are exchanged. In terms of the elliptic modulus of the curve, this is precisely an $S$-duality transformation, 
	\be
	\label{sdual}
	\tau_D=-{1\over \tau}. 
	\ee	
The dual theory will play a r\^ole in the discussion of the boundary conditions for the holomorphic ambiguity. 
 
As we explained in the previous section, quantum corrections to $t$, $t_D$ can be computed explicitly by using the all-orders WKB method. 
For example, in order to compute the free energy at next-to-leading order, we note that
\be
{\rd F_1 \over \rd \nu}= t_D^{(1)}\left(\xi_0(\nu) \right) - \pi \ri \tau\left(\xi_0(\nu) \right) t^{(1)} \left(\xi_0(\nu) \right). 
\ee
After integration, one finds that, up to an additive constant, 
\be
\label{f1-cub}
F_1(\nu)=-\frac{1}{24} \log ( \nu )+\frac{77 \nu }{32}+\frac{13937 \nu
   ^2}{256}+\frac{1717793 \nu ^3}{1024}+\frac{240109947 \nu
   ^4}{4096}+\CO\left(\nu ^5\right).
   \ee
This turns out to agree with the expansion of
\be
\label{f1dis}
F_1(\nu)=-{1\over 24} \log \Delta (\xi), 
\ee
where $\Delta(\xi)$ is the discriminant (\ref{disc-cub}). This is precisely what one expects for the subleading correction to NS free energy in 
supersymmetric gauge theories and topological strings \cite{kw,hk}. This suggests that the higher order corrections to the quantum free energy 
can be computed with the holomorphic anomaly. 

In order to study the holomorphic anomaly equations and its direct integration for the curve (\ref{yx-cubic}), 
we need to write the theory in terms of modular forms. The shortest path 
to this consists in mapping the curve (\ref{yx-cubic}) to the Seiberg--Witten (SW) form \cite{sw}
\be
\label{swcurve}
y^2= (x^2-1)(x-u).  
\ee
This is achieved by a simple linear change of variables. We choose the map in such a way that 
the conifold point $\xi=0$ of (\ref{yx-cubic}) corresponds to $u=1$ (the monopole point of the SW curve). One finds that $\xi$ is related to $u$ by 
\be
 \xi= {1\over 108 g^2} \left\{ 1+ {u (u^2-9) \over (u^2+3)^{3/2}}\right\}. 
 \ee
 We now recall that $u$ can be written in terms of theta functions of the modulus $\tau$ in (\ref{tau-k}), as follows (see for example \cite{hk}), 
 \be
 u(\tau)= {\vartheta_2^4(\tau) +  \vartheta_3^4(\tau) \over \vartheta_4^4(\tau)}. 
 \ee
 It will be useful to express the modular forms in terms of the generators
 \be
 \ba
 K_2\left(q\right) &:= \vartheta_3^4\left(\tau\right) + \vartheta_4^4\left(\tau\right),\\
 K_4\left(q\right)&:=\vartheta_2^8\left(\tau\right), 
 \ea
 \ee
 as in \cite{kmr}. These are appropriate for the monodromy group $\Gamma(2)$ characterizing the SW curve (\ref{swcurve}). By using all these ingredients, 
 it is easy to derive the following explicit formula for the Yukawa coupling, 
 \begin{equation}
 \label{yuk-cub}
		Y = \frac{16 \sqrt{2} \left(K_2^2+3 K_4\right)^{9/4}}{\left(K_2^2-K_4\right)^2 K_4}.
\end{equation}
We can now write the holomorphic anomaly equation in the normalization appropriate for our theory. It reads
\begin{equation}
	\frac{\partial F_n}{\partial \widehat E_2} = -\frac{Y^2}{24} \:\: \sum_{r=1}^{n-1} D_\tau F_r \: D_\tau F_{n-r}.
	\end{equation}
As usual in the direct integration approach, the most delicate point is to find a good ansazt for the holomorphic ambiguity. As in \cite{kmr}, 
we will work in the ring of non-holomorphic modular forms generated by $K_2$, $K_4$ and $\widehat E_2$. Note that this 
ring is closed under the action of the Maass derivative, since 
\be
\label{mod-ders}
\ba	D_{\tau} K_2 &= \frac{1}{12} \left(-K_2^2+2 \widehat E_2 K_2+3 K_4\right),\\
		D_{\tau} K_4 &= \frac{1}{3} \left(K_2+\widehat E_2\right) K_4,\\
		D_{\tau} \widehat E_2 &= \frac{1}{48} \left(-K_2^2-3 K_4+4 \widehat E_2^2\right).
\ea
\ee
The appropriate parametrization for the ambiguity turns out to be 
\be
\label{amb-cub}
f_n(\tau)=\frac{\left(K_2^2+3 K_4\right)^{\frac{3 (n+1)}{2}}}
		{\left(K_2^2-K_4\right){}^{4 (n-1)} K_4^{2 (n-1)}} \sum_{i=0}^{  \left\lfloor \frac{9n-15}{2} \right\rfloor } a_i K_2^{9n-15-2i}K_4^i.
	\end{equation}
The coefficients $a_i$ are fixed by requiring the expansion of $F_n(\nu)$ near $\nu=0$ to be of the 
form (\ref{sr}), where the singular part is given by (\ref{singfn}), (\ref{coni-bound}), with $k_n=1$ for all $n\ge 0$. This is the standard gap condition of \cite{hk,hkk}, specialized to the 
NS limit of the free energies. In these quantum-mechanical models, this boundary condition can be justified as follows. The singular part of the quantum $B$-period 
\be
{\rd F^{\rm s} \over \rd \nu}= \sum_{n \ge 0} {\rd F^{\rm s}_n \over \rd \nu}, 
\ee
where we have set $\hbar=1$, is precisely the asymptotic expansion of the function 
\be
\log \left[ {\Gamma \left( {1\over 2} + \nu \right) \over {\sqrt{2 \pi}} 8^\nu}\right]
\ee
at $\nu \rightarrow \infty$ (note that the constant appearing in the first line of (\ref{singfn}) has the value $\nu_0=8$ in this model). 
But the term inside the brackets is the (inverse) prefactor in (\ref{fnu}). Therefore, the structure of the $B$-period, as determined 
in \cite{ddp,al-cas}, {\it implies} the standard gap condition. Note in particular that the function $A(\nu)$ is given by the regular part of the quantum $B$-period 

The boundary condition (\ref{singfn}) 
does not fix all the coefficients in the holomorphic ambiguity. We will 
require in addition that the {\it dual} function $F_n^D(\tau)$, obtained by an $S$-duality transformation 
(\ref{sdual}) of $F_n(\tau)$, satisfies the same gap condition when expanded in $\nu_D=\tilde t_D$. 
In this problem, it turns out that the general dual expansion involves half-integer powers of $\tilde t_D$, 
which must also be required to vanish. This provides $6n-6$ conditions to fulfill, so the resulting system for the coefficients $a_i$ is overdetermined. 
We have found experimentally that it has a unique solution for $2\le n\le 30$. For example, for $F_2$, we find in this way, after taking 
the holomorphic limit, 
\be
F_2 =- {K_2^2 \left(K_2^2-9 K_4\right)^2 \left(K_2^2+3 K_4\right)^{5/2} \over 108 \left(K_2^2-K_4\right)^4 K_4^2} E_2 -{79 K_2 \left(K_2^2-9 K_4\right) \left(K_2^2+3 K_4\right)^{9/2} \over 1080 \left(K_2^2-K_4\right)^4 K_4^2}. 
\ee
The first term is fully determined by the anomaly equation, while the second one is the holomorphic ambiguity. By expanding around $\nu=0$, we obtain 
\be
F_2(\nu)=-\frac{7}{5760 \nu^2}+\frac{101479}{30720}+\frac{43147783
   \nu}{122880}+\frac{1769452671 \nu^2}{65536}+\frac{1185129116647
   \nu^3}{655360}+\CO\left(\nu^4\right).
\ee
The dependence on $g$ is easily restored by the rescaling, 
\be
\label{rec-res}
\nu \rightarrow g^2 \nu, \qquad F_n\rightarrow (g^2)^{2n-2} F_n. 
\ee
It is easy to generate expressions for the very first $F_n$'s in terms of $E_2$, $K_2$ and $K_4$, although their explicit form is too long to be copied here. For their 
expansions around $\nu=0$, we find for example, 
\be
\ba
F_3(\nu)&=\frac{31}{161280 \nu ^4}+\frac{791845439}{655360}+\frac{724731745353 \nu
   }{2621440}+\frac{157755456235861 \nu ^2}{4194304}+\CO\left(\nu ^3\right), \\
 F_4(\nu)&=-\frac{127}{1290240 \nu ^6}+\frac{1430873478800591}{1006632960}+\frac{1367769982181464281 \nu }{2684354560}+\CO\left(\nu ^2\right).
   \ea
   \ee
These expansions reproduce the known results for the coefficients (\ref{c-cub}) as listed for example in \cite{casares-tesis} (see also \cite{gt} for similar data). This 
 validates our claim that the all-orders WKB periods of the cubic oscillator are governed by the refined holomorphic anomaly equations. Note in particular that, by using 
 the explicit expressions for $F_n(\tau)$ in terms of modular forms, the equations (\ref{afr}) and (\ref{mod-ders}), and the explicit expression for the Yukawa coupling (\ref{yuk-cub}), one 
 can write the function $A(\nu)$ (together with the singular part) as an infinite formal power series in $\hbar$ whose coefficients are modular forms. 
  
Interestingly, the above ansatz for the holomorphic ambiguity generates a non-trivial constant term $\mu_n$ in the regular part of the free energies 
$F^{\rm r}_n(\nu)$, which has the form 
\be
F^{\rm r}_n (\nu)=   \mu_n + \CO(\nu).
\ee
The presence of this constant term is at first glance surprising, since the WKB method only determines the {\it derivative} of $F_n(\nu)$ w.r.t. $\nu$, so 
the constant $\mu_n$ is not fixed by quantum special geometry. What is its meaning? 

The PNP relation for the cubic oscillator, first found by \'Alvarez and Casares in \cite{al-cas-2}, reads
\be
{\rd \xi \over \rd \nu}=-{15  g^2 \over 2} \left( \nu + g^2 {\partial A \over \partial g^2} \right). 
\ee
By using (\ref{afr}) we see that, up to an integration constant independent of $\nu$,
\be
\label{xiF}
\xi(\nu)= -{15 g^2 \nu^2 \over 4}- {15 \over 2} g^4 {\partial F^{\rm r}(\nu) \over \partial g^2}. 
\ee
The quantum mirror map at $\nu=0$ is given by 
\be
\xi(0)=-{7 g^2\over 16}-{101479 g^6 \over 2048}-{2375536317 g^{10}\over 65536}+ \cdots, 
\ee
as calculated in for example \cite{gt}. It is then natural to choose a constant term $\mu_n$ in $F_n$, $n\ge 2$, 
in such a way that (\ref{xiF}) is satisfied also for the $\nu$-independent part, 
i.e. such that 
\be
\xi(0)+{7 g^2\over 16}= -15  g^4 \sum_{n \ge 2} (n-1) \mu_n (g^2)^{2n-3}. 
\ee
This is precisely the value of $\mu_n$ which is obtained with the ansazt (\ref{amb-cub}) for the holomorphic ambiguity, as one can verify from the 
explicit solution of the $F_n$. It is uniquely determined by the ansatz and the gap conditions. Therefore, the anomaly equation makes it possible to define an integrated 
form for the PNP relation, involving the quantum free energy, and with the right choice of integration constant. In fact, after including in addition the right constant for $n=1$, and 
using the full free energy, one can write down the simpler relation, 
\be
\label{xiFt}
\xi(\nu)=-{3g^2\over 4} -{15  g^4\over 2} {\partial F \over \partial g^2}, 
\ee
where we reintroduce $g$ with (\ref{rec-res}) before taking the derivative. As opposed to (\ref{xiF}), this includes the $\nu$-independent term.

Let us make one final remark. One can verify from the explicit solutions for the quantum free energies that 
\be
F_n^D(\tau)= F_n (\tau), 
\ee
i.e. the cubic oscillator is a self-dual model. Physically, this reflects the fact that inverting the cubic oscillator we obtain a completely equivalent system.

\subsection{The double well and the quartic oscillator}

\begin{figure}[tb]
\begin{center}
\resizebox{75mm}{!}{\includegraphics{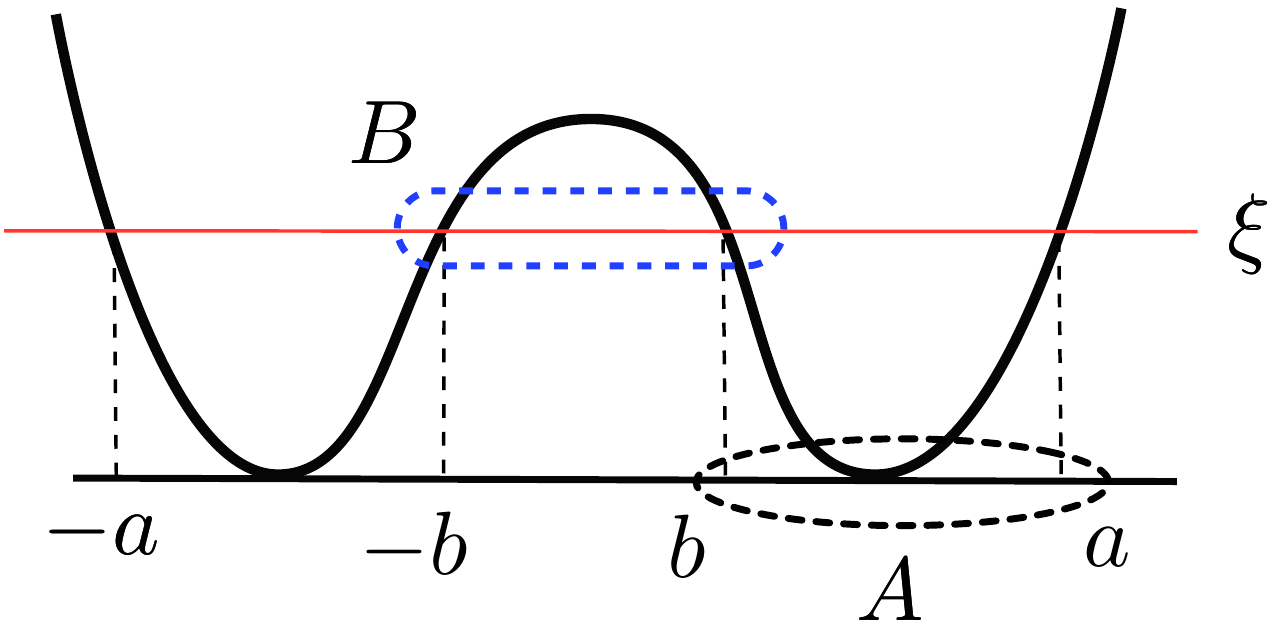}}
\end{center}
  \caption{The symmetric double well. 
}
\label{dw-fig}
\end{figure}
The symmetric double well describes a quantum-mechanical particle in the potential 
\be
V(x)= {x^2\over 2}\left(1+ g x\right)^2. 
\ee
This can be made symmetric w.r.t. $x=0$ by shifting $x\rightarrow x-1/(2g)$. In this symmetric realization, it has the form shown in \figref{dw-fig}. The symmetric double well is 
a textbook example for the importance of non-perturbative effects in Quantum Mechanics (see for example \cite{mm-book}). In perturbation theory, the model has 
two degenerate minima, but exponentially small quantum effects, due to instanton tunneling, lift  the degeneracy. The exact quantization condition for this model was first 
conjectured in \cite{zj} by using multi-instanton calculus\footnote{We should however note that a structurally identical, exact quantization condition for the pure quartic oscillator was already 
written down explicitly by Voros in \cite{voros}.}. It can be derived in the context of the all-orders WKB method by using the 
exact connection formula of Voros--Silverstone (a related derivation appears in \cite{ddp}). Another derivation, based on the uniform WKB method, was presented by \'Alvarez in \cite{alvarez}. 
The resulting condition reads, 
\be
\label{dw-qc}
  1+  \exp\left( \pm {\ri \over \hbar}  \oint_A P(x) \rd x \right) = \epsilon \ri \exp\left( {\ri \over 2\hbar} 
  \oint_B P(x) \rd x\right). 
  \ee
Here, $\epsilon=\pm1$ refers to the parity of the state, while the $\pm $ sign in the l.h.s. refers to the choice of lateral Borel resummation. The function appearing in the r.h.s.,
\be
\label{fnu-dw}
f(\nu)= \exp\left( {\ri \over 2\hbar}  \oint_B P(x) \rd x\right),
\ee
has the structure:
\be
f(\nu)={{\sqrt{2 \pi}} \over \Gamma\left({1\over 2}+ \nu\right)} \left( {2 \over g^2} \right)^\nu \re^{-A(\nu)/2}, 
\ee
where $A(\nu)$, defined in (\ref{afr}), is also the function appearing in the quantization condition of \cite{zj,zjj1} (we have set $\hbar=1$ for simplicity.) This structural 
result can be obtained in the framework of \cite{ddp}. It was also derived by \'Alvarez in \cite{alvarez} by using the uniform 
WKB method\footnote{In order to make contact with \cite{alvarez}, we note that: i) the function $f(\nu)$, as defined in \cite{alvarez}, 
includes the factor 
of $\ri$ appearing in the r.h.s. of (\ref{dw-qc}), ii) our 
variable $\nu$ has to be identified with his $\nu/{\sqrt{2}}$, and iii) his $g$ has to be set to ${\sqrt{2}}g^2$.}. 
As in the cubic case, the function $A(\nu)$ has the expansion 
\be
\label{a-dw}
A(\nu)= {1\over 3g^2}+ \sum_{k \ge 1} c^{(k)}(\nu) g^{2k}, 
\ee
where the coefficients $c^{(k)}(\nu)$ are polynomials in $\nu$ with rational coefficients. They can be computed by performing the WKB integrals, as in \cite{zjj1,zjj2}, 
or by using asymptotic matching in the uniform WKB method of \cite{alvarez} (as in the case of the cubic oscillator, this method has the advantage of producing the function $A(\nu)$ directly as a function 
of the quantum period, while the standard WKB technique, as applied in \cite{zjj1,zjj2}, gives $A(\nu)$ as a function of $\xi$). One obtains, at the very first orders (see also 
\cite{du})
\be
\label{c-dw}
\ba
c^{(1)}(\nu)&= {19 \over 12} + 17 \nu^2,\\
c^{(2)}(\nu)&= {153\nu\over 4} + 125 \nu^3,\\
c^{(3)}(\nu)&={22709 \over 576} + {23405 \nu^2 \over 24}+ {17815 \nu^4 \over 12}. 
\ea
\ee

Let us now study this model using special geometry and the refined holomorphic anomaly equations. 
The elliptic curve for the double well can be taken to be
\be
\label{yx-dw}
y^2=(x^2-a^2)(b^2-x^2), 
\ee
where we have rescaled the variables as in (\ref{rescaling}) to set $g=1$. We have
\be
a^2={1\over 4} \left( 1+ {\sqrt{32 \xi}}\right), \qquad b^2={1\over 4} \left( 1- {\sqrt{32 \xi}}\right). 
\ee
 The classical periods are given by the integrals,  
\be
t= {1\over \pi} \int_b^a y(x) \rd x, \qquad t_D= -2 \ri \int_{-b}^b y(x) \rd x. 
\ee
The Picard--Fuchs equation reads in this case
\begin{equation}
		\left[6 + \xi\left(32 \xi-1\right) \frac{\partial^2}{\partial \xi^2} \right] \Pi= 0. 
	\end{equation}
The classical $t$-period can be expanded around $\xi=0$ as in (\ref{tex}), where $c_1=1$ and the $c_n$ satisfy now the recursion
\be
c_n = \frac{2 (4 n-7) (4 n-5)}{(n-1) n} c_{n-1}, \qquad n \ge 2. 
\ee
As in the cubic case, it is also possible to write $t$ and $t_D$ in terms of elliptic integrals of the first and second kind (see also section 4.1 of \cite{kmr}, 
where this calculation was performed in the context of the cubic Hermitian matrix model.) The results are as follows:
\be
\label{dw-expers}
\ba
t&=  {a b^2 \over 3 \pi k'^{2}}\left[ (1+ k'^{2}) E(k) - 2 k'^{2} K(k) \right],\\
t_D&={4 a b^2 \over 3 (k')^2} \left[ (1+k'^{2}) E(k')- k^2 K(k') \right],
\ea
\ee
where the elliptic modulus is 
\be
\label{k-dw}
k^2=1-{b^2 \over a^2}. 
\ee
By using the above results, one derives the following expansion of the classical periods around $\xi=0$:
\be
\ba
t=&  \xi +3 \xi ^2+35 \xi ^3+\frac{1155 \xi ^4}{2}+\frac{45045 \xi ^5}{4}+\cdots\\
t_D=& \frac{1}{3}+2 \xi  \left(\log \left(\frac{\xi }{2}\right)-1\right)+ \xi ^2 \left(6 \log \left(\frac{\xi	}{2}\right)+17 \right)+\cdots. 
\ea
\ee
The prepotential $F_0(t)$ is then given by 
\be
\label{prep-dw}
F_0(t)= \frac{t}{3}+t^2 \left(\log \left(\frac{t}{2}\right)-\frac{3}{2}\right)+\frac{17 t^3}{3}+\frac{125 t^4}{4}+\frac{3563 t^5}{12}+\frac{29183
				t^6}{8}+\cdots. 
\ee
The $\tau$ parameter of the curve will be given again by $\ri K'/K$, where the elliptic modulus is (\ref{k-dw}), but in this case we have 
\be
\tau= {1\over 4\pi \ri} {\partial t_D \over \partial t}. 
\ee

The discriminant of the curve (\ref{yx-dw}) is given by 
\be
\label{dis-dw}
\Delta(\xi)=\xi^2 (1-32 \xi). 
\ee
This leads to two singular points. The first one is the standard conifold point at $\xi=0$, while the dual conifold point is 
\be
\xi={1\over 32}. 
\ee
This is the value of $\xi$ in which the energy equals the height of the barrier. 
As in the cubic theory, we define a dual energy as
\be
\label{xid-xi-dw}
\xi_D={1\over 32} - \xi, 
\ee
and $\xi_D=0$ is the dual conifold point. The classical period $t_D$ 
can be expanded around $\xi_D=0$ as
\be
t_D= {\sqrt{2}} \pi \tilde t_D, \qquad \tilde t_D= 2 \xi_D+6 \xi_D^2 + 70 \xi_D^3+\cdots 
\ee
Note that $\tilde t_D/2$ has the same expansion in terms of $\xi_D$, than $t$ in terms of $\xi$. Physically, the dual situation corresponds to inverting the 
double-well and exchanging the perturbative and the tunneling cycles, up to an overall factor. The resulting model is then the inverted quartic oscillator. As 
we will see, the quantum free energy of this inverted quartic oscillator can be obtained by considering the S-dual theory to the double-well (i.e. by implementing the 
S-dual transformation (\ref{sdual})) and identifying $\tilde t_D$ with the quantum $A$-period. The dual theory will also play a r\^ole in the 
discussion of the boundary conditions for the holomorphic ambiguity. 
 
We can now study the quantum corrections to $t$, $t_D$ obtained with the all-orders WKB method. One finds that the first correction to the 
quantum free energy is given again by the formula (\ref{f1dis}). 
Higher order corrections can be obtained by direct integration of the holomorphic anomaly equations. In this 
case, life is even simpler since we can borrow the results from \cite{kmr}. The Yukawa coupling was obtained in terms of modular forms in \cite{kmr} and it reads, 
 \begin{equation}
		Y =   \frac{128 \sqrt{2} K_2^{5/2}}{K_4 \left( K_2^2 -K_4\right)}.
\end{equation}
The holomorphic anomaly equation is now given by 
\begin{equation}
	\frac{\partial F_n}{\partial \widehat E_2} = -\frac{Y^2}{192} \:\: \sum_{r=1}^{n-1} D_\tau F_r \: D_\tau F_{n-r},
	\end{equation}
where we have chosen the appropriate normalization. The holomorphic ambiguity turns out to be parametrized as, 
\be
\label{amb-dw}
f_n(\tau)=\left(\frac{1}{\left(K_2^2-K_4\right) K_4}\right)^{2 (n-1)} K_2^{\rho_n} \sum_{i=0}^{4(n-1)-\rho_n/2} a_i K_2^{8(n-1)-\rho_n-2i}K_4^{i},
	\end{equation}
	where 
	\be
	\rho_n=\begin{cases} 3n-3, &\text{if $n$ is odd}, \\
	3n-2, &\text{if $n$ is even}. \end{cases}
	\ee
The coefficients $a_i$ are fixed by the boundary conditions. Since the log of (\ref{fnu-dw}) gives half of the quantum $B$-period, 
the expansion of $F_n(\nu)$ near $\nu=0$ should be of the 
form (\ref{coni-bound}), with $k_n=2$ for all $n\ge 0$ (in this example, $\nu_0=2$). Let us now define the dual free energy by 
\be
\label{dual-fn}
F_n^D(\tau)= (-2)^{1-g} F_n (\tau) \Bigl|_{\tau \rightarrow -1/\tau}. 
\ee
We will require this function to have a pole of order $2n-2$ when expanded as a function of $\tilde t_D$ around $\tilde t_D=0$. 
Putting everything together, we find $4n-5$ constraints and only $4n-3-\rho_n/2$ unknowns, so the system is overdetermined. However, 
it has a solution in all cases we have checked, up to $n=40$. 
The coefficient of the singularity in the dual expansion is fixed by these conditions and it is equal to (\ref{d-coni-bound}), with $k_n^D=1$. 
Using this procedure we find for example, 
\be
\ba
F_2 &= -\frac{2  \left(2 K_2^2-3 K_4\right){}^2 K_2^3}{27 \left(K_2^2-K_4\right){}^2 K_4^2} \widehat{E}_2 -\frac{2 \left(158 K_2^5-330 K_4 K_2^3+135 K_4^2
	K_2\right) K_2^3}{135 \left(K_2^2-K_4\right){}^2 K_4^2}, 
   \\
 F_3&=\frac{32  \left(2 K_2^2-3 K_4\right){}^3 K_2^7}{2187 \left(K_2^2-K_4\right){}^4 K_4^4} \widehat{E}_2^3+
	 \frac{16  \left(2 K_2^2-3 K_4\right){}^2 \left(22 K_2^4-39 K_4 K_2^2+27 K_4^2\right) K_2^6}{729 \left(K_2^2-K_4\right){}^4 K_4^4} \widehat{E}_2^2 + \\
	 &+  \frac{8  \left(2 K_2^2-3 K_4\right)
 	\left(7604 K_2^6-23088 K_4 K_2^4+23625 K_4^2 K_2^2-6345 K_4^3\right) K_2^7}{3645 \left(K_2^2-K_4\right){}^4 K_4^4}\widehat{E}_2 + f_3(\tau), 
   \ea
   \ee
   where the holomorphic ambiguity at order $n=3$ reads
   \be
   \ba
   f_3(\tau)&=\frac{4 K_2^6 }{76545  	\left(K_2^2-K_4\right){}^4 K_4^4}
   \left(7384904 K_2^{10}-31999716 K_4 K_2^8+ 53857062 K_4^2 K_2^6-\right.\\
   &\qquad \qquad \qquad \left. -43355655 K_4^3 K_2^4+16924950 K_4^4 K_2^2-1927233 K_4^5\right). 
   \ea
\ee
By expanding around $\nu=0$, we obtain 
\be
\ba
F_2(\nu)=&-\frac{7}{2880 \nu^2}+\frac{131}{192}+\frac{22709 \nu}{576}+\frac{217663
   \nu^2}{128}+\frac{61936297 \nu^3}{960}+\frac{581912191 \nu^4}{256}+\CO\left(\nu^5\right), \\
   F_3(\nu)=&\frac{31}{80640 \nu ^4}+\frac{10483}{256}+\frac{20182631 \nu }{3840}+\frac{628163783 \nu
   ^2}{1536}+\frac{178000040363 \nu ^3}{7168}+O\left(\nu ^4\right). 
   \ea
\ee
The dependence on $g$ is restored again by (\ref{rec-res}). The explicit expressions for $F_4$ and the higher $n$ 
free energies become too long to be copied here, but we can write 
the first terms of their expansion around $\nu=0$. We have, for example, 
\be
 F_4(\nu)=  -\frac{127}{645120 \nu ^6}+\frac{193438987}{24576}+\frac{553607616973 \nu
   }{344064}+\frac{12106775869037 \nu ^2}{65536}+\CO\left(\nu ^3\right).     \ee
   As in the cubic model, the $A(\nu)$ function appearing in (\ref{a-dw}) is related to the regular part of the quantum free energy by the equation 
   (\ref{afr}) (the regular part is defined as in the cubic oscillator, by removing logarithmic terms and poles in the expansion of the $F_n$). The expansions 
   of the $F_n$ obtained above agree with previous results in quantum mechanics, as listed in for example 
   \cite{alvarez} (see also \cite{du}). 
   
 The PNP relation for the double well, found by \'Alvarez in \cite{alvarez}, can be written as   
\be
{\rd \xi \over \rd \nu}=-3  g^2  \left(2 \nu + g^2 {\partial A \over \partial g^2} \right). 
\ee
After integration, we find that 
\be
\label{xiF-dw}
\xi(\nu)= -{3 g^2 \nu^2 }- 3 g^4 {\partial F^{\rm r}(\nu) \over \partial g^2},
\ee
up to a $\nu$ independent term.
The quantum mirror map at $\nu=0$ is given by 
\be
\xi(0)=-{ g^2\over 4}-{131 g^6 \over 32}- {31449 g^{10}\over 64}-{580316961 g^{14} \over 4096}-\cdots. 
\ee
This can be easily calculated for example with the BenderWu package of \cite{su}. The constant term $\mu_n$ in $F_n$ should be such that 
(\ref{xiF-dw}) is satisfied for the $\nu$-independent part, i.e. such that 
\be
\xi(0)+{g^2\over 4} =-3g^4 \sum_{n \ge 2} (2n-2) \mu_n (g^2)^{2n-3}. 
\ee
As in the cubic oscillator, this is precisely the value of $\mu_n$ obtained with the holomorphic anomaly equations. One can use the full quantum free energy 
to write the PNP relation as in (\ref{xiFt}), 
 \be
 \label{xifdw}
\xi(\nu)=-{g^2 \over 2} -3 g^4 {\partial F \over \partial g^2}, 
\ee
where again we reintroduce $g$ with (\ref{rec-res}) before the derivative.

Finally, we note that results for the quantum free energies of the (unstable) quartic oscillator with potential 
\be
V(x)= {x^2 \over 2} - g x^4
\ee
can be obtained from the results for the double well, after an $S$-duality transformation. 
The quantum free energies of the quartic oscillator are precisely the dual free energies in (\ref{dual-fn}), 
while $\tilde t_D$ is the quantum $A$-period. One obtains, for example,
\be
\ba
F_2^D(\tilde t_D)&= -\frac{7}{5760 \tilde t_D^2}+{513 \over 512} + \frac{305141 \tilde t_D}{9216}+\frac{3105983
  \tilde t_D^2}{4096}+\frac{912774217 \tilde t_D^3}{61440}+\cdots, \\
  F_3^D(\tilde t_D)&= \frac{31}{161280
			\tilde t_D^4}+\frac{485523}{8192}+\frac{1056412343
			\tilde t_D}{245760}+\frac{34978331399
			\tilde t_D^2}{196608}+\cdots
			\ea
  \ee
This and other results agree with the calculations using uniform WKB in \cite{al-cas, casares-tesis}. They also agree with the calculation of the $A$ function in \cite{jsz}, after 
expressing it in terms of the quantum period as in \cite{gt}. 

\sectiono{Application: the large order behavior of the WKB expansion}

In the all-orders WKB method, the energy is given in terms of the quantum number $m$ by the formal power series (\ref{qmm}) in $\hbar^2$, 
where $\nu$ is given by (\ref{num}). After reexpanding this series in powers of $g$, we recover the perturbative Rayleigh--Schr\"odinger 
series for the $m$-th energy level. The large order behavior of this perturbative series is governed by instanton corrections, 
as first noted by Bender and Wu in their seminal paper \cite{bw} (see \cite{mm-book} for a textbook presentation). However, one can ask a slightly different 
question: what is the large order behavior of (\ref{qmm}), i.e. the behavior 
of the coefficients $\xi_n(\nu)$ when $n$ is large? This is a series in $\hbar^2$, not in the coupling constant, and in addition its coefficients are non-trivial functions of $\nu$. 
As we will see, its behavior is 
different from the one of the standard perturbative series. Since the direct integration of the holomorphic anomaly equation, combined with the PNP relations, easily generates 
many terms in the expansion (\ref{qmm}), one can test in detail the predictions for its large order behavior. 

We will now determine the large order behavior of (\ref{qmm}) by using standard tools in resurgent analysis 
(see for example \cite{mm-rev,mm-book} for a simple introduction, and \cite{msw,mm-effects} for very similar 
applications). We first note that this series can be 
extended to a {\it trans-series} involving an exponentially small parameter, of the form
\be
\label{xits}
\Xi(\nu)=\sum_{\ell=0}^\infty \re^{-\ell  \CA(\nu)/\hbar} \xi^{(\ell)}(\nu), 
\ee
where
\be
\xi^{(\ell)}(\nu)= \sum_{r \ge 0}  \xi^{(\ell)}_r (\nu) \hbar^r.
\ee
Here, $\CA(\nu)$ is the instanton action, and $\xi^{(0)}(\nu)$ is the original asymptotic series $\xi(\nu)$ in (\ref{qmm}). Note that $\xi^{(0)}_r (\nu)=0$ for 
odd $r$. The trans-series (\ref{xits}) 
is obtained by solving the appropriate quantization condition. We will focus on the case of the double well, 
so we need the trans-series solution to (\ref{dw-qc}). We choose the $-$ sign in this equation, and we write the result in the form 
\be
\label{eqnu}
\ri f(\widehat \nu)= 1+ \re^{-2\pi \ri \widehat \nu/\hbar}. 
\ee
We have denoted the argument of this equation by $\widehat \nu$ to emphasize the fact that it now has non-perturbative corrections and is no longer 
given by (\ref{num}). We have also reabsorbed the parity $\epsilon$ inside the function $f(\widehat \nu)$. We now sketch the derivation of the trans-series $\Xi(\nu)$, 
following a similar derivation in \cite{alvarez}. The solution to (\ref{eqnu}) is of the form 
\be
\widehat \nu= \nu + \Delta \nu, 
\ee
where $\nu$ is given by (\ref{num}) and $\Delta \nu$ is non-perturbative in $\hbar$. It satisfies the equation 
\be
\label{delnu}
\Delta \nu = {\ri \hbar \over 2 \pi} \log \left(1- \ri f \left(\nu + \Delta \nu \right) \right). 
\ee
We note that, from the defining expression in (\ref{fnu-dw}), one has, 
\be
f(\nu) =\exp \left( -{\CA (\nu) \over \hbar}  \right) \left\{ 1- \hbar {\partial F_1 \over \partial \nu} +\cdots \right\}, 
\ee
where
 \be
 \CA(\nu) ={1\over 2} {\partial F_0 \over \partial \nu}={t_D(\nu) \over 2} 
 \ee
is the instanton action, and $t_D(\nu)$ is given in (\ref{dw-expers}). We now solve for $\Delta \nu$ as an instanton expansion, 
\be
\Delta \nu= \sum_{k \ge 1} \Delta \nu ^{(k)}, 
\ee
where 
\be
\Delta \nu ^{(k)} \propto \re^{- k \CA(\nu)/\hbar}. 
\ee
From (\ref{delnu}) one obtains, 
\be
\ba
\Delta \nu ^{(1)}&= {\hbar \over 2 \pi} f(\nu), \\
\Delta \nu ^{(2)}&= {\ri \hbar \over 4 \pi} f^2(\nu) + \left( {\hbar \over 2\pi}\right)^2 f'(\nu) f(\nu). 
\ea
\ee
We finally obtain, up to instanton number two, 
\be
\Xi (\nu) = \xi( \widehat \nu)= \xi(\nu)+ {\partial \xi \over \partial \nu} \Delta \nu ^{(1)}+ {\partial \xi \over \partial \nu} \Delta \nu ^{(2)}+ {1\over 2} {\partial^2 \xi \over \partial \nu^2} \left( \Delta \nu ^{(1)}\right)^2+\cdots, 
\ee
and we conclude that
\be
\label{trans-series}
\ba
 \re^{-\CA(\nu)/\hbar } \xi^{(1)}(\nu)&={\hbar \over 2 \pi} f(\nu)  {\partial \xi \over \partial \nu},\\
  \re^{-2 \CA(\nu)/\hbar } \xi^{(2)}(\nu)&={\ri \hbar \over 4 \pi} f^2(\nu)  {\partial \xi \over \partial \nu}+{\hbar^2 \over 8 \pi^2} {\partial \over \partial \nu} \left(  f^2(\nu)  {\partial \xi \over \partial \nu}\right). 
  \ea
  \ee
As is well-known, the large order behavior is controlled by the imaginary part of the trans-series, and the 
 first contribution appears at the two-instanton level \cite{bpz}. The imaginary part of the first two coefficients in $\xi^{(2)}(\nu)$ is given by,  
 \be\label{imxis}
{\rm Im} \, \xi^{(2)}_0(\nu)= {\hbar  \over 4 \pi} {\partial \xi_0\over \partial \nu}, \qquad 
 {\rm Im}\,  \xi^{(2)}_1(\nu)=- {\hbar \over 4 \pi} {\partial \xi_0\over \partial \nu} {\partial F_1 \over \partial \nu}.
 \ee
The higher order coefficients ${\rm Im}\, \xi^{(2)}_r(\nu)$ with $r\ge 2$ can be computed in a straightforward way. 
Standard arguments (see for example \cite{mm-book,msw}) 
tell us that the large order behavior of the WKB series $\xi_n(\nu)$ is of the form\footnote{In this section $A(\nu)$ denotes 
the instanton action, and it should not be confused with the function defined in (\ref{afr}).},  
\be
\label{lo-xin}
\xi_n (\nu) \sim {1 \over \pi} (A (\nu))^{-2n-b}  \Gamma(2n+b)\, \mu_1(\nu) \left[1 + {\mu_2 (\nu) A(\nu) \over 2n} + \cdots \right], \quad n \gg 1, 
\ee
where the values of $A (\nu)$, $b$ and the $\mu_n(\nu)$ can be obtained from the leading behavior of the imaginary part of 
the trans-series, i.e. from $\re^{-2 \CA(\nu)} \xi^{(2)}(\nu)$. 
One finds, 
\be
\label{ex-values}
A(\nu)= 2 \CA(\nu)=t_D(\nu), \qquad b=-1, \qquad 
\mu_1(\nu)= -{1 \over 2 \pi} {\partial \xi_0\over \partial \nu}, \qquad 
\mu_2(\nu)=- {\partial F_1 \over \partial \nu}. 
\ee
The factor of $2$ and the minus sign in $\mu_1$ are due to the fact that the relevant non-perturbative quantity is the difference 
between the two lateral resummations, and this introduces a factor of $-2$.

It is interesting to note that the WKB expansion has the $(2n)!$ growth typical of a string theory or 
of the $1/N$ expansion \cite{shenker,mm-rev}. This can be regarded as yet another consequence of the fact that this expansion 
is governed by the refined holomorphic anomaly equations of topological string theory, as we argued in this paper. 
In contrast, the standard perturbative expansion in the coupling constant 
grows like $n!$. In addition, the WKB series and its asymptotics depend on $\nu$, which plays the r\^ole of a 
't Hooft parameter (as already suggested in \cite{krefl,krefl2,bd}).

\begin{figure}[tb]
\begin{center}
\resizebox{115mm}{!}{\includegraphics{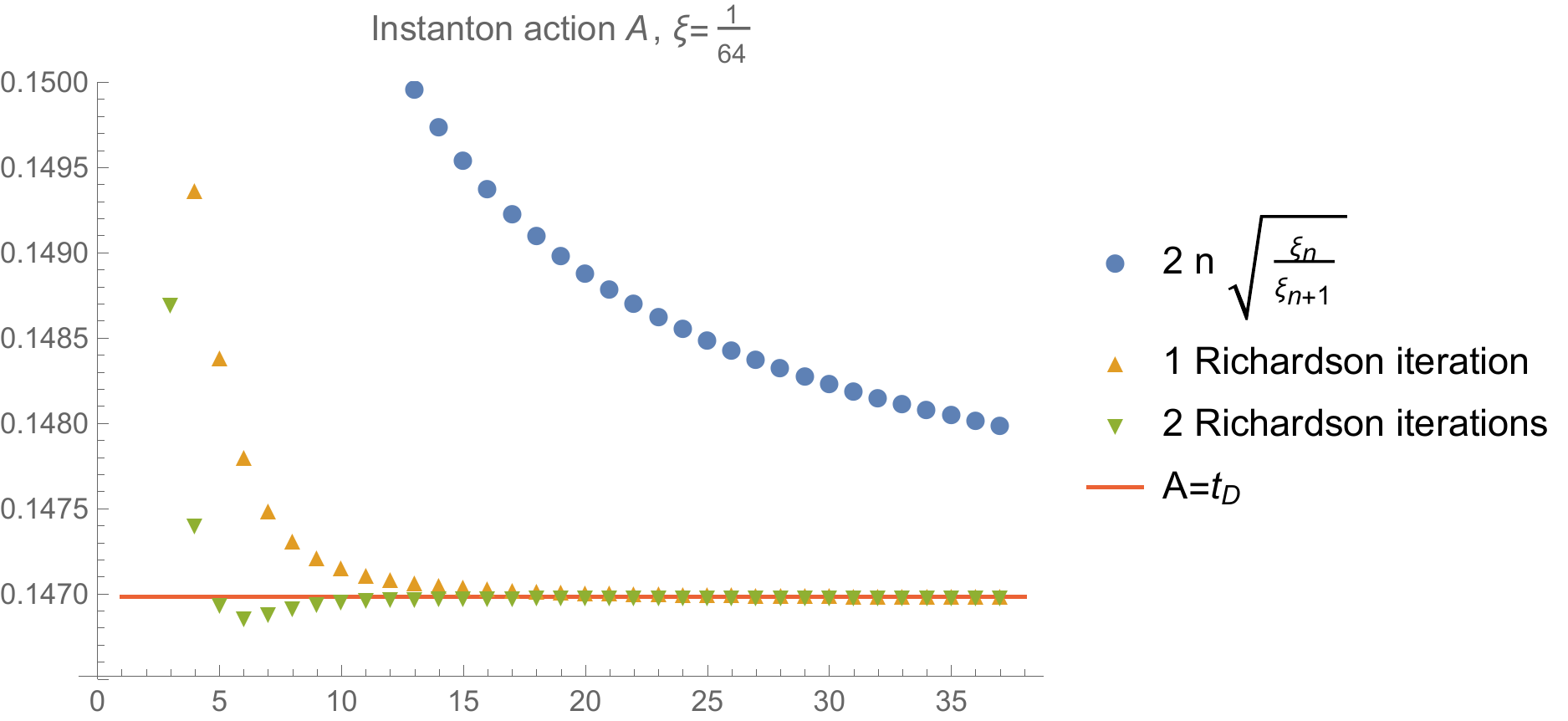}}
\end{center}
  \caption{The sequence (\ref{richainst}) for the double well and $\xi=1/64$, together with two Richardson transforms. 
}
\label{a-test}
\end{figure}

The resurgent prediction (\ref{lo-xin}) for the large order behavior can be tested in detail by studying the 
sequence $\xi_n(\nu)$ for high enough $n$. From this original sequence, one can consider various auxiliary sequences to extract the coefficients. 
For example, to extract $A$, we consider the sequence
\be\label{richainst}
Q_n=2n {\sqrt{ \xi_{n+1} \over \xi_n}}=A+\CO\left({1\over n} \right).
\ee
\begin{figure}[tb]
\begin{center}
\resizebox{125mm}{!}{\includegraphics{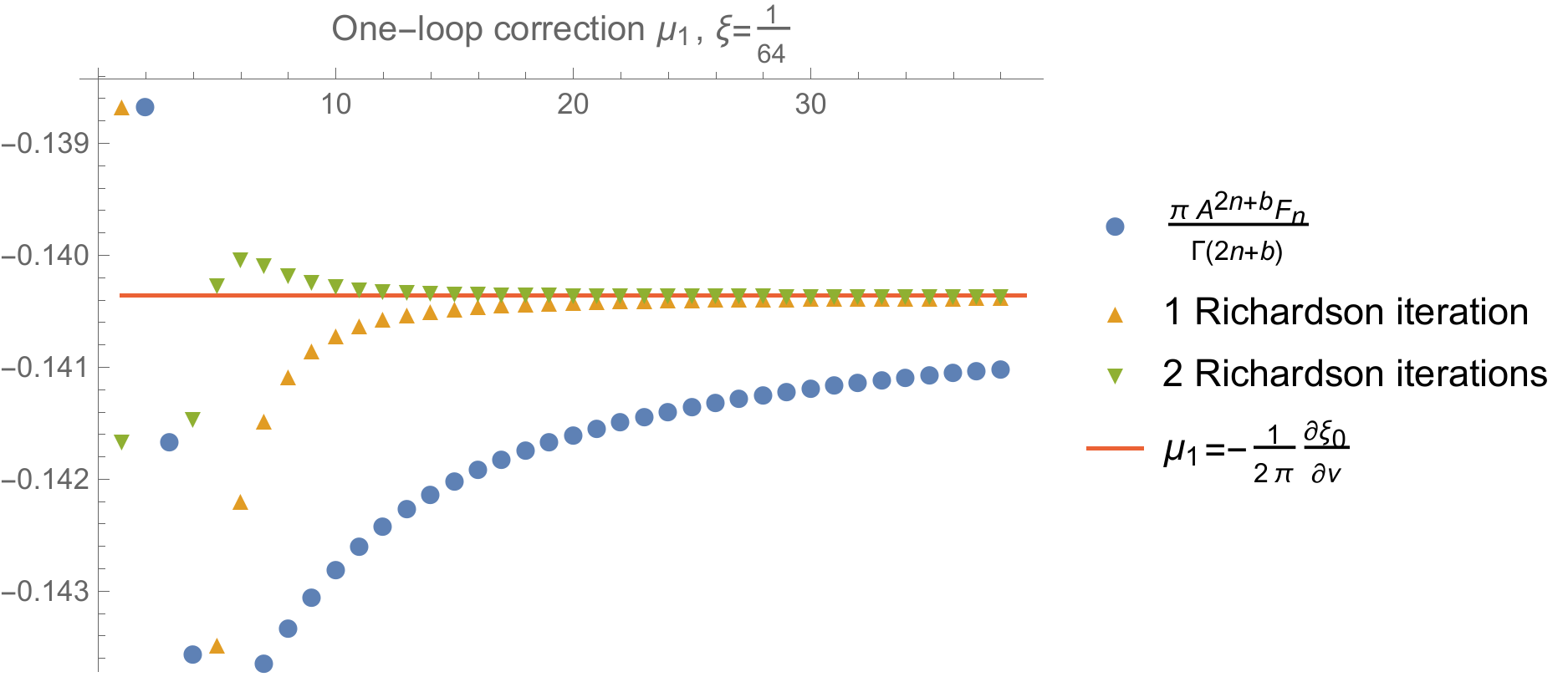}}
\end{center}
  \caption{The sequence (\ref{richc0}) for the double well and $\xi=1/64$, together with two Richardson transforms. 
}
\label{mu1-test}
\end{figure}
The parameter $b$ follows from the sequence
\be
n\left(A^2{\xi_{n+1}\over 4n^2\xi_n}-1\right)-{1\over 2}=b+\CO\left({1\over n}\right).
\ee
\noindent
The coefficients $\mu_1$ and $\mu_2$ are obtained from the sequences
\be\label{richc0}
{\pi A^{2n+b}\xi_n\over \Gamma(2n+b)}=\mu_1\left(1+{\mu_2 A\over 2n}+\CO\left({1\over n^2}\right)\right)
\ee
\noindent
and
\be\label{richc1}
{2n\over A}\left({\pi A^{2n+b}\xi_n\over \mu_1\Gamma(2n+b)}-1 \right)=\mu_2+\CO\left({1\over n}\right), 
\ee
respectively. The convergence of these sequences can be accelerated with Richardson transforms, as explained in for example \cite{msw}. We obtain in this way 
very good numerical approximations to the expected values (\ref{ex-values}), which are themselves functions of $\nu$ (or, equivalently, of the 
modulus $\xi$, which is identified with $\xi_0(\nu)$). 

\begin{figure}[tb]
\begin{center}
\resizebox{125mm}{!}{\includegraphics{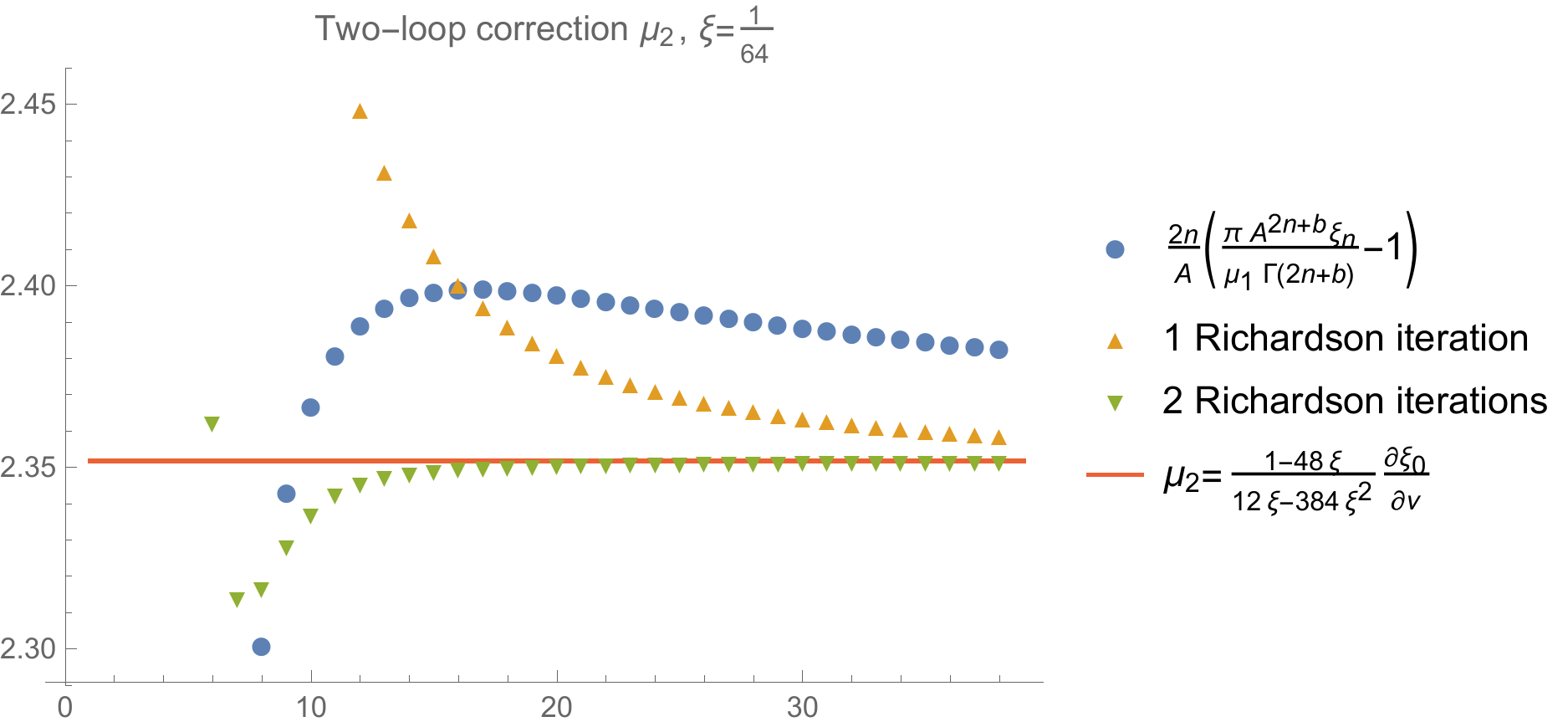}}
\end{center}
  \caption{The sequence (\ref{richc1}) and two Richardson transforms for the double well and $\xi=1/64$. 
}
\label{mu2-test}
\end{figure}

Let us give an example of the procedure. By combining the expression (\ref{xifdw}) and the results of direct integration, 
we generated the first forty terms in the sequence $\xi_n(\nu)$. For the value $\xi=1/64$ (which corresponds to $\nu=0.01654...$), we show 
in \figref{a-test} the sequence (\ref{richainst}), together with two Richardson 
transformations, as well as the prediction $t_D$. Numerically, the best estimate for the instanton action is given by 
\be
A^{\rm num}=0.14698331391354... 
\ee
The prediction from resurgent analysis is 
\be
t_D=0.1469833139135404...
\ee
Similarly, we show in \figref{mu1-test} and \figref{mu2-test} the sequences (\ref{richc0}) and (\ref{richc1}), respectively, for $\xi=1/64$, together with their Richardson transforms, and we compare them to the 
theoretical predictions. Numerically, the best estimates are 
\be
\mu^{\rm num}_1=-0.1403587257682..., \qquad \mu^{\rm num}_2= 2.3517330226...,
\ee
while the expected values are 
\be
\mu_1=-0.140358725768206..., \qquad \mu_2=2.351733022617...
\ee
In calculating $\mu_2$, we have used that 
\be
{\partial F_1 \over \partial \nu}= -{1\over 24 \Delta} {\rd \Delta \over \rd \xi}{\partial \xi_0 \over \partial \nu}, 
\ee
where $\Delta(\xi)$ is the discriminant (\ref{dis-dw}). 
The numerical estimates obtained above are excellent, if we take into account that we have used only forty terms in the series. They are comparable
to similar results in the instanton analysis of large $N$ matrix models \cite{msw,mm-effects}. We have checked the large order predictions for many other 
values of $\xi$ (including complex ones), and in all cases we find an excellent agreement with the numerical results. 

Finally, let us note that the WKB series for the energy (\ref{qmm}) in the double well inherits the well-known lack of Borel summability of the perturbative series, since 
the instanton action $t_D$ is positive for any $0\le \xi<1/32$ inside the radius of convergence. It would be interesting to study how to recover the exact result for the energies 
by using the Borel--\'Ecalle resummation of the WKB (trans-)series, instead of the resummation of the perturbative series performed in \cite{zjj1,zjj2}.

\sectiono{Conclusions and outlook}

In this paper we have argued that the refined holomorphic anomaly equations of topological string theory govern the 
all-orders WKB periods of general one-dimensional quantum mechanical systems. In the case of quantum mechanical models related 
to $\CN=2$ supersymmetric gauge theories or to topological string theory in the NS limit, like the (modified) Mathieu equation, this is expected. 
However, we claim that this connection is more general and holds for oscillators which do not have a known realization in terms of gauge theories, 
strings or matrix models, like the cubic oscillator. 

Our results seem to indicate that the holomorphic anomaly equations, in the NS limit, are just a consequence of the all orders WKB method of \cite{dunham}. 
It would be very interesting to prove that, indeed, the equations (\ref{rha-ns}) follow from the recursion (\ref{y-recur}). Note that a similar situation occurs 
for matrix models in the $1/N$ expansion. In principle, the genus $g$ free energies of matrix models are holomorphic, but in 
the context of the topological recursion \cite{eo}, one can introduce 
a non-holomorphic dependence through the Bergmann kernel of the curve, which is a basic building block of the recursion. It is then relatively simple 
to show that the resulting non-holomorphic free energies, computed by a non-holomorphic extension of the topological recursion, 
satisfy the standard holomorphic anomaly equation of \cite{bcov}, as shown in \cite{emo}. In the case of the all-orders WKB periods, the results derived in \cite{alba} for topological strings provide 
a strategy to derive (\ref{rha-ns}). One important assumption needed in a derivation along the lines of \cite{alba} is that the quantum periods 
can be obtained from the classical periods by acting with differential operators depending on the energy (which is the modulus of this problem). This assumption 
is known to hold in many cases. It might be possible to derive it systematically from the recursion (\ref{y-recur}). 

Our results also require a clarification concerning the literature on quantum curves (see for example \cite{d-mulase} for a review). 
As it is well known, given a spectral curve, the topological recursion of \cite{eo} leads to a formal WKB-like wavefunction $\psi_{\rm top}$. 
It is often claimed that $\psi_{\rm top}$ is 
annihilated by an appropriate or ``natural" quantization of the underlying spectral curve, so that it provides a formal WKB solution 
to the operator problem defined by the quantum version of the curve. 
In the cases considered in this paper, for example, the ``natural" quantization simply involves
the promotion of $y$, $x$ to canonically conjugate Heisenberg operators
\be
\label{quanta}
y \rightarrow \mathsf{y}, \qquad x \rightarrow \mathsf{x}, \qquad [\mathsf{x}, \mathsf{y}]=\ri \hbar,  
\ee
so that we recover the standard Schr\"odinger operator (\ref{schrodinger}). It has been recently proved that this claim is true when the spectral 
curve has genus zero \cite{beynard}. However, the situation for curves of higher 
genus is more subtle. One hint that the topological recursion cannot lead straightforwardly to the WKB result is that 
the $F_g$'s produced by the topological recursion, as applied for example to the curves studied in this paper, are simply {\it not} the functions $F_n$ produced 
by the WKB method\footnote{In the case of the double-well, this is easily seen by comparing the results in \cite{kmr} to the results in section 3.2 above.}. 
This is only to be expected: the former corresponds to the unrefined topological string theory, 
while the latter corresponds to the NS limit of the refined string. In line with this observation, a 
recent study of the Weierstrass curve, which has genus one, indicates that $\psi_{\rm top}$ is not annihilated by the quantum curve obtained 
by the quantization scheme (\ref{quanta}) \cite{bouchard}, in other words, it is not a formal WKB wavefunction associated to the natural 
quantization of the curve.

The connection unveiled in this paper between the standard WKB method and the holomorphic anomaly equations is potentially 
interesting to further clarify the quantum mechanical meaning of the general equations (\ref{rha}). 
It was famously suggested by Witten in \cite{witten} that the standard topological string partition function should be interpreted 
as a wavefunction. The holomorphic anomaly equation then implements the independence of this wavefunction w.r.t. the choice of polarization. 
If our point of view is correct, there might be a precise connection between the holomorphic anomaly equations and a Schr\"odinger-like 
equation governing the partition function.  

There are many concrete extensions and generalizations 
of our approach to similar problems and settings. One could study the holomorphic anomaly equations for other one-dimensional models, like the 
Fokker--Planck potential or the cosine potential. Although we have focused for simplicity on models leading to curves of genus one, we expect that higher genus curves will be 
also governed by the refined holomorphic anomaly equations. In the case of curves of genus two, one could use already the technology developed in \cite{kpsw}. 
Notice as well that, by combining the results found in this paper with the results of \cite{mp}, it is possible to compute
 the full $1/N$ expansion for the ground state energy of Matrix Quantum Mechanics, in terms of modular forms. This is an appealing laboratory to explore 
 the structure and properties of the $1/N$ expansion at all orders. The analysis of the large order behavior in section 4 can be clearly extended to the cubic and the quartic 
 oscillators, and as we already mentioned in that section, it would be interesting to understand the Borel--\'Ecalle resummation of the WKB expansion in all the examples considered in this paper. 
 Finally, although in the examples considered in this paper we can compute the relevant trans-series from the explicit quantization conditions, it would be interesting to study the general 
 trans-series by extending the refined holomorphic anomaly equations to the non-perturbative sector, as it was done in \cite{cesv} for the standard topological string.  

\acknowledgments
We would like to thank Gabriel \'Alvarez, Vincent Bouchard, Alba Grassi, Albrecht Klemm and Ricardo Schiappa for discussions 
and correspondence. This research was supported in part by the Fonds National Suisse, subsidies 200021-156995 and 200020-141329, 
and by the Swiss-NSF grant NCCR 51NF40-141869 ``The Mathematics of Physics'' (SwissMAP).


\begin{thebibliography}{99}
\bibliographystyle{plain}

\bibitem{adkmv} 
M.~Aganagic, R.~Dijkgraaf, A.~Klemm, M.~Mari\~no and C.~Vafa, ``Topological strings and integrable hierarchies,''
  Commun.\ Math.\ Phys.\  {\bf 261}, 451 (2006) [hep-th/0312085].


\bibitem{ns}
 N.~A.~Nekrasov and S.~L.~Shatashvili, ``Quantization of Integrable Systems and Four Dimensional Gauge Theories,''
  arXiv:0908.4052 [hep-th].

\bibitem{mm-stms}
  M.~Mari\~no, ``Spectral Theory and Mirror Symmetry,''
  arXiv:1506.07757 [math-ph].
  
  \bibitem{nekrasov}
N.~A.~Nekrasov, ``Seiberg-Witten prepotential from instanton counting,''
  Adv.\ Theor.\ Math.\ Phys.\  {\bf 7}, no. 5, 831 (2003) [hep-th/0206161].
  
  
\bibitem{hm}
Y.~Hatsuda and M.~Mari\~no, ``Exact quantization conditions for the relativistic Toda lattice,''
  JHEP {\bf 1605}, 133 (2016) [arXiv:1511.02860 [hep-th]].
  
   


 \bibitem{mirmor}
 A.~Mironov and A.~Morozov, ``Nekrasov Functions and Exact Bohr-Sommerfeld Integrals,''
  JHEP {\bf 1004}, 040 (2010) [arXiv:0910.5670 [hep-th]].
  
\bibitem{acdkv}
M.~Aganagic, M.~C.~N.~Cheng, R.~Dijkgraaf, D.~Krefl and C.~Vafa, ``Quantum Geometry of Refined Topological Strings,''
  JHEP {\bf 1211}, 019 (2012)
  [arXiv:1105.0630 [hep-th]].

\bibitem{he}
 W.~He, ``Combinatorial approach to Mathieu and Lam\'e equations,''
  J.\ Math.\ Phys.\  {\bf 56}, no. 7, 072302 (2015) [arXiv:1108.0300 [math-ph]].
  
  \bibitem{krefl}
D.~Krefl,``Non-Perturbative Quantum Geometry,''
  JHEP {\bf 1402}, 084 (2014) [arXiv:1311.0584 [hep-th]].
  
\bibitem{krefl2}   D.~Krefl, ``Non-Perturbative Quantum Geometry II,''
  JHEP {\bf 1412}, 118 (2014) [arXiv:1410.7116 [hep-th]].

  \bibitem{bd}
 G.~Basar and G.~V.~Dunne, ``Resurgence and the Nekrasov-Shatashvili limit: connecting weak and 
 strong coupling in the Mathieu and Lam\'e systems,''
  JHEP {\bf 1502}, 160 (2015) [arXiv:1501.05671 [hep-th]].
  
    
\bibitem{kpt}
 A.~K.~Kashani-Poor and J.~Troost, ``Pure $ \mathcal{N}=2 $ super Yang-Mills and exact WKB,''
  JHEP {\bf 1508}, 160 (2015) [arXiv:1504.08324 [hep-th]].
  
\bibitem{ashok}  S.~K.~Ashok, D.~P.~Jatkar, R.~R.~John, M.~Raman and J.~Troost, ``Exact WKB analysis of $ \mathcal{N} $ = 2 gauge theories,''
  JHEP {\bf 1607}, 115 (2016) [arXiv:1604.05520 [hep-th]].
  
  \bibitem{kw} 
D.~Krefl and J.~Walcher, ``Extended Holomorphic Anomaly in Gauge Theory,''
  Lett.\ Math.\ Phys.\  {\bf 95}, 67 (2011) [arXiv:1007.0263 [hep-th]].
  
  \bibitem{hk}
M.~x.~Huang and A.~Klemm, ``Direct integration for general $\Omega$ backgrounds,''
  Adv.\ Theor.\ Math.\ Phys.\  {\bf 16}, no. 3, 805 (2012)
  [arXiv:1009.1126 [hep-th]].

  
\bibitem{bcov}
 M.~Bershadsky, S.~Cecotti, H.~Ooguri and C.~Vafa, ``Kodaira-Spencer theory of gravity and exact results for quantum string amplitudes,''
  Commun.\ Math.\ Phys.\  {\bf 165}, 311 (1994) [hep-th/9309140].

 \bibitem{hk-ha} 
 M.~x.~Huang and A.~Klemm, ``Holomorphic Anomaly in Gauge Theories and Matrix Models,''
  JHEP {\bf 0709}, 054 (2007) [hep-th/0605195].

\bibitem{hkr}
 B.~Haghighat, A.~Klemm and M.~Rauch, ``Integrability of the holomorphic anomaly equations,''
  JHEP {\bf 0810}, 097 (2008) [arXiv:0809.1674 [hep-th]].
 
 \bibitem{dv}
 R.~Dijkgraaf and C.~Vafa, ``Matrix models, topological strings, and supersymmetric gauge theories,''
  Nucl.\ Phys.\ B {\bf 644}, 3 (2002) [hep-th/0206255].
  
  \bibitem{emo}
 B.~Eynard, M.~Mari\~no and N.~Orantin, ``Holomorphic anomaly and matrix models,''
  JHEP {\bf 0706}, 058 (2007) [hep-th/0702110 [HEP-TH]].

\bibitem{kmr}
 A.~Klemm, M.~Mari\~no and M.~Rauch, ``Direct Integration and Non-Perturbative Effects in Matrix Models,''
  JHEP {\bf 1010}, 004 (2010) [arXiv:1002.3846 [hep-th]].
  
  \bibitem{dmp}
   N.~Drukker, M.~Mari\~no and P.~Putrov, ``From weak to strong coupling in ABJM theory,''
  Commun.\ Math.\ Phys.\  {\bf 306}, 511 (2011) [arXiv:1007.3837 [hep-th]].

  
\bibitem{voros}
A. Voros, {\it Spectre de l'\'equation de Schr\"odinger et m\'ethode BKW}, Publications Math\'ematiques d'Orsay, 1981, in \url{http://sites.mathdoc.fr/PMO/PDF/V_VOROS-167.pdf}. 

\bibitem{voros-return}
A. Voros, ``The return of the quartic oscillator. The complex WKB method," Annales de l'I.H.P. {\bf 39}, 211 (1983).

\bibitem{zj}
 J.~Zinn-Justin, ``Multi - Instanton Contributions in Quantum Mechanics. 2.,'' Nucl.\ Phys.\ B {\bf 218}, 333 (1983). 

\bibitem{dunham}  J. L. Dunham,``The Wentzel-Brillouin-Kramers method of solving the wave equation," Phys. Rev. {\bf 41}, 713 (1932).
  
  \bibitem{shcp}
 H. J. Silverstone, J. G. Harris, J. C'zek and J. Paldus, ``Asymptotics of high-order perturbation theory for the one-dimensional anharmonic oscillator by quasisemiclassical methods," 
 Phys. Rev. A {\bf 32}, 1965 (1985).
 
  \bibitem{al-cas}
G. \'Alvarez and C. Casares, ``Uniform asymptotic and JWKB 
expansions for anharmonic oscillators," J. Phys. A {\bf 33}, 2499 (2000).  

\bibitem{al-cas-2}
G. \'Alvarez and C. Casares, ``Exponentially small corrections in the asymptotic expansion of the eigenvalues 
of the cubic anharmonic oscillator," J. Phys. A {\bf 33}, 5171 (2000).

\bibitem{ahs}
G. \'Alvarez, C. J. Howls, and H. J. Silverstone, ``Anharmonic oscillator discontinuity formulae up to 
second-exponentially-small order," J. Phys. A {\bf 35}, 4003 (2002).

\bibitem{alvarez}
G. \'Alvarez, ``Langer--Cherry derivation of the multi-instanton expansion for the symmetric double well," J. Math. Phys. {\bf 45}, 3095 (2004).

\bibitem{gkmw}
 T.~W.~Grimm, A.~Klemm, M.~Mari\~no and M.~Weiss,``Direct Integration of the Topological String,''
  JHEP {\bf 0708}, 058 (2007) [hep-th/0702187 [HEP-TH]].

  
  \bibitem{huang-beta}
   M.~x.~Huang, ``Dijkgraaf-Vafa conjecture and $\beta$-deformed matrix models,''
  JHEP {\bf 1307}, 173 (2013) [arXiv:1305.1103 [hep-th]].


  
  
\bibitem{zjj1}
J.~Zinn-Justin and U.~D.~Jentschura, ``Multi-instantons and exact results I: Conjectures, WKB expansions, and instanton interactions,''
  Annals Phys.\  {\bf 313}, 197 (2004) [quant-ph/0501136].

\bibitem{zjj2}
 J.~Zinn-Justin and U.~D.~Jentschura, ``Multi-instantons and exact results II: Specific cases, higher-order effects, and numerical calculations,''
  Annals Phys.\  {\bf 313}, 269 (2004) [quant-ph/0501137].
  
  \bibitem{jsz}
U.~D.~Jentschura, A.~Surzhykov and J.~Zinn-Justin, ``Multi-instantons and exact results. III: Unification of even and odd anharmonic oscillators,''
  Annals Phys.\  {\bf 325}, 1135 (2010).
  
   \bibitem{bpz}
 E.~Br\'ezin, G.~Parisi and J.~Zinn-Justin, ``Perturbation Theory at Large Orders for Potential with Degenerate Minima,''
  Phys.\ Rev.\ D {\bf 16}, 408 (1977). 
  
  \bibitem{shenker}
  S.H.~Shenker, ``The strength of nonperturbative effects in string theory," in O.~\'Alvarez, E.~Marinari and P.~Windey (eds.), {\it Random Surfaces and Quantum Gravity}, Plenum, New York, 1992.
  
  
\bibitem{mm-rev}
  M.~Mari\~no, ``Lectures on non-perturbative effects in large $N$ gauge theories, matrix models and strings,''
  Fortsch.\ Phys.\  {\bf 62}, 455 (2014) [arXiv:1206.6272 [hep-th]].

\bibitem{mm-book} 
M. Mari\~no, {\it Instantons and large $N$. An introduction to non-perturbative methods in Quantum Field Theory}, Cambridge University Press, 2015.   

\bibitem{gp2} A. Galindo and P. Pascual, {\it Quantum Mechanics}, volume 2, Springer--Verlag, 1991. 

\bibitem{ags}
G. \'Alvarez, S. Graffi and H. J. Silverstone, ``Transition from classical mechanics to quantum mechanics: $x^4$ perturbed harmonic oscillator," 
Phys. Rev. {\bf A 38}, 1687 (1988).

\bibitem{alvarez3}
G. \'Alvarez, ``Quantum mechanics as classical mechanics plus quantum corrections: the cubic anharmonic oscillator," J. Phys. A {\bf 22}, 617 (1989).


 
 \bibitem{hkk}
  M.~x.~Huang, A.~K.~Kashani-Poor and A.~Klemm, ``The $\Omega$ deformed B-model for rigid $\mathcal{N}=2$ theories,''
  Annales Henri Poincare {\bf 14}, 425 (2013) [arXiv:1109.5728 [hep-th]].
  
    \bibitem{huang}
 M.~x.~Huang, ``On Gauge Theory and Topological String in Nekrasov-Shatashvili Limit,''
  JHEP {\bf 1206}, 152 (2012)
  [arXiv:1205.3652 [hep-th]].
  
\bibitem{ckk}
 J.~Choi, S.~Katz and A.~Klemm, ``The refined BPS index from stable pair invariants,''
  Commun.\ Math.\ Phys.\  {\bf 328}, 903 (2014) [arXiv:1210.4403 [hep-th]].
 
    \bibitem{kpsw}
 A.~Klemm, M.~Poretschkin, T.~Schimannek and M.~Westerholt-Raum, ``Direct Integration for Mirror 
 Curves of Genus Two and an Almost Meromorphic Siegel Modular Form,''
  arXiv:1502.00557 [hep-th]. 

  

  

  
  \bibitem{hoe}
 N. Hoe {\it et al.}, ``Stark effect of hydrogenic ions," Phy. Rev. A {\bf 25}, 891 (1982). 
 
   
  
\bibitem{du}
G.~V.~Dunne and M.~\"Unsal, ``Uniform WKB, Multi-instantons, and Resurgent Trans-Series,''
  Phys.\ Rev.\ D {\bf 89}, no. 10, 105009 (2014) [arXiv:1401.5202 [hep-th]].
  
\bibitem{gorsky}
A.~Gorsky and A.~Milekhin, ``RG-Whitham dynamics and complex Hamiltonian systems,''
  Nucl.\ Phys.\ B {\bf 895}, 33 (2015) [arXiv:1408.0425 [hep-th]].


\bibitem{alba}
   A.~Grassi, ``Spectral determinants and quantum theta functions,''
  J.\ Phys.\ A {\bf 49}, no. 50, 505401 (2016) [arXiv:1604.06786 [hep-th]].
  
  \bibitem{bender}
  R. Yaris, J. Bendler, R. Lovett, C. M. Bender, P. A. Fedders, ``Resonance calculations for arbitrary potentials," Phys. Rev. A {\bf 18}, 1816 (1978). 
  
\bibitem{alvarez-cub}
G. \'Alvarez, ``Coupling-constant behavior of the resonances of the cubic anharmonic oscillator," Phys. Rev. A {\bf 37}, 4079 (1988).


  
   \bibitem{gg} E. Caliceti, S. Graffi and M. Maioli, ``Perturbation theory of odd anharmonic oscillators," 
  Comm. Math. Phys. {\bf 75}, 51 (1980).
  
\bibitem{caliceti}
E.~Caliceti, M.~Meyer-Hermann, P.~Ribeca, A.~Surzhykov and U.~D.~Jentschura, 
``From Useful Algorithms for Slowly Convergent Series to Physical Predictions Based on Divergent Perturbative Expansions,'' 
Phys. Rep. {\bf 446} (2007) 1 [arXiv:0707.1596 [physics.comp-ph]].

  
\bibitem{silverstone} 
H. J. Silverstone, ``JWKB connection-formula problem revisited via Borel summation," Phys. Rev. Lett. {\bf 55}, 2523 (1985).

  
\bibitem{ddp}
E. Delabaere, H. Dillinger and F. Pham, ``Exact semiclassical expansions for 
one-dimensional quantum oscillators," J. Math. Phys. {\bf 38}, 6126 (1997).



\bibitem{casares-tesis}
C. Casares, {\it Desarrollos asint\'oticos y f\'ormulas de conexi\'on para la ecuaci\'on de Schr\"odinger con potencial 
polin\'omico}, Ph.D. thesis, 2003. 



\bibitem{gt}
 I.~Gahramanov and K.~Tezgin, ``Remark on the Dunne-\"Unsal relation in exact semiclassics,''
  Phys.\ Rev.\ D {\bf 93}, no. 6, 065037 (2016) [arXiv:1512.08466 [hep-th]].


 
 \bibitem{sw}
  N.~Seiberg and E.~Witten, ``Electric - magnetic duality, monopole condensation, and confinement in N=2 supersymmetric Yang-Mills theory,''
  Nucl.\ Phys.\ B {\bf 426}, 19 (1994); Erratum: [Nucl.\ Phys.\ B {\bf 430}, 485 (1994)] [hep-th/9407087].

\bibitem{su}
T.~Sulejmanpasic and M.~\"Unsal, ``Aspects of Perturbation theory in Quantum Mechanics: The BenderWu Mathematica package,''
  arXiv:1608.08256 [hep-th].

  


\bibitem{bw}
 C.~M.~Bender and T.~T.~Wu, ``Anharmonic oscillator. 2: A Study of perturbation theory in large order,''
  Phys.\ Rev.\ D {\bf 7}, 1620 (1973).
 
   
 \bibitem{msw}
 M.~Mari\~no, R.~Schiappa and M.~Weiss, ``Nonperturbative Effects and the Large-Order Behavior of Matrix Models and Topological Strings,''
  Commun.\ Num.\ Theor.\ Phys.\  {\bf 2}, 349 (2008) [arXiv:0711.1954 [hep-th]].

 
 \bibitem{mm-effects}
 M.~Mari\~no, ``Nonperturbative effects and nonperturbative definitions in matrix models and topological strings,''
  JHEP {\bf 0812}, 114 (2008) [arXiv:0805.3033 [hep-th]].



  
  


\bibitem{eo}
B.~Eynard and N.~Orantin, ``Invariants of algebraic curves and topological expansion,''
  Commun.\ Num.\ Theor.\ Phys.\  {\bf 1}, 347 (2007) [math-ph/0702045].

\bibitem{d-mulase}
 O.~Dumitrescu and M.~Mulase, ``Lectures on the topological recursion for Higgs bundles and quantum curves,''
  arXiv:1509.09007 [math.AG].
\bibitem{beynard}
 V.~Bouchard and B.~Eynard, ``Reconstructing WKB from topological recursion,''
  arXiv:1606.04498 [math-ph].
  
  \bibitem{bouchard}
  V.~Bouchard, N.~K.~Chidambaram and T.~Dauphinee, ``Quantizing Weierstrass,''
  arXiv:1610.00225 [math-ph].
  
  \bibitem{ggu}
 A.~Grassi and J.~Gu, ``BPS relations from spectral problems and blowup equations,''
  arXiv:1609.05914 [hep-th].

\bibitem{witten}
E.~Witten, ``Quantum background independence in string theory,''
  Salamfest 1993:0257-275
  [hep-th/9306122].

  
  \bibitem{mp}

M.~Mari\~no and P.~Putrov, ``Multi-instantons in large $N$ Matrix Quantum Mechanics,''
  arXiv:0911.3076 [hep-th].

 \bibitem{cesv} 
 R.~Couso-Santamar\'\i a, J.~D.~Edelstein, R.~Schiappa and M.~Vonk, ``Resurgent Transseries and the Holomorphic Anomaly,''
  Annales Henri Poincare {\bf 17}, no. 2, 331 (2016) [arXiv:1308.1695 [hep-th]]; 
  ``Resurgent Transseries and the Holomorphic Anomaly: Nonperturbative 
  Closed Strings in Local ${\mathbb{C}\mathbb{P}^2}$,'' Commun.\ Math.\ Phys.\  {\bf 338}, no. 1, 285 (2015) [arXiv:1407.4821 [hep-th]].

\end{thebibliography}
\end{document}